\documentclass[final,3p,times,twocolumn,number,sort&compress,superscriptaddress]{elsarticle}

\usepackage{graphicx}

\usepackage{color}

\usepackage{amssymb}

\usepackage{lineno}

\journal{Journal of Magnetism and Magnetic Materials}

\begin{document}

\begin{frontmatter}


\title{Electric polarization induced by phase separation in magnetically ordered and
paramagnetic states of RMn$_2$O$_5$ (R = Gd, Bi)}

\author[pti]{B.Kh.~Khannanov\corref{cor1}}
\ead{boris.khannanov@gmail.com}

\author[pti]{V.A.~Sanina\corref{cor2}}
\ead{sanina@mail.ioffe.ru}

\author[pti]{E.I.~Golovenchits}
\ead{e.golovenchits@mail.ioffe.ru}

\author[pti]{M.P.~Scheglov}
\ead{m.scheglov@mail.ioffe.ru}

\cortext[cor1]{Corresponding author}
\cortext[cor2]{Principal corresponding author}

\address[pti]{A.F. Ioffe Physical Technical Institute RAS, 26 Politekhnicheskaya,
194021, St. Petersburg, Russia}

\begin{abstract}
The electric polarization hysteresis loops and remanent polarization were revealed
in multiferroics RMn$_2$O$_5$ with $R = Gd$ and $Bi$ at wide temperature interval from 5 K up to 330 K.
Until recently, the long-range ferroelectric order having an exchange-striction magnetic nature
had been observed in RMn$_2$O$_5$ only at low temperatures ($T\leq T_C = 30$ -- 35 K).
We believe that the polarization we observed was caused by the frozen superparaelectric state which was
formed by the restricted polar domains resulting from phase separation and charge carriers self-organization.
At some sufficiently high temperatures $T\gg T_C$ the frozen superparaelectric state was destroyed,
and the conventional superparaelectric state occurred. This happened when the potential barriers
of the restricted polar domain reorientations become equal to the kinetic energy of the itinerant electrons (leakage).
The hysteresis loops were measured by the so-called PUND method which allowed us to correctly
subtract the contribution of conductivity from the measured polarization. The correlations between
properties of the phase separation domains and polarization were revealed and studied.
The high-temperature polarization also had a magnetic nature and was controlled by the magnetic
field because the double exchange between pairs of Mn ions with different valences (Mn$^{3+}$ and Mn$^{4+}$)
in RMn$_2$O$_5$ was the basic interaction resulting in phase separation.
\end{abstract}

\begin{keyword}
multiferroic \sep polarization hysteresis loop \sep phase separation \sep charge carrier self-organization

\PACS 75.85.+t \sep 77.22.Ej \sep 77.80.Dj
\end{keyword}

\end{frontmatter}

\section{Introduction}
\label{intro}

\noindent Manganites RMn$_2$O$_5$ (R are rare earth ions, Y and Bi) are typical representatives
of multiferroics in which ferroelectric (with $T_C\simeq 30 -35$ K) and magnetic (with $T_N\simeq 35 - 40$ K)
orderings coexist ~\cite{nature2003,HurNature2004,KatsuraPRL2005,MostovoyPRL2006,Noda2008}.
It has been assumed until now that RMn$_2$O$_5$ has a centrosymmetric sp. gr. {\it Pbam} which forbids
the existence of electric polarization.
The ferroelectric ordering along the $b$ axis at low temperatures is induced by charge and magnetic orders
which break the central symmetry of the crystal due to the exchange-striction mechanism  ~\cite{Brink2008}.
It has recently been shown ~\cite{PRL2015} that the real RMn$_2$O$_5$ symmetry at room temperature
is a noncentral symmetry which means that the electric polarization should exist at room temperature.
The authors could not give preference to one of two possible space groups: {\it P2}
(allowing polarization along the $c$ axis) and {\it Pm} (allowing polarization in the $ab$ plane)
on the basis of structural studies. By assuming from the physical considerations
that the {\it homogeneous} electric polarization could exist at all temperatures only in one direction
(i.e., along the $b$ axis), they preferred the sp. gr. {\it Pm}.
The authors ~\cite {PRL2015} did not discuss the nature of this polarization.
In ~\cite{JETPLett2016} it was reported that the hysteresis loops of electric polarization
and remanent polarization were detected in GdMn$_2$O$_5$ in a wide temperature range 5 -- 330 K.
We interpreted this polar state as a frozen superparaelectric state induced by {\it local polar domains}
formed in the main crystal matrix due to phase separation and charge carriers self-organization.
Theoretically, the frozen superparaelectric
state was considered in the system of isolated ferroelectric
nanoscale domains in a dielectric matrix ~\cite{Glinchuk}.

This paper presents a comparative study of the electric polarization hysteresis loops
which were revealed in two crystals of the RMn$_2$O$_5$ family with $R = Gd$ and $Bi$ in temperature
interval from 5 K up to 330 K. There are significant structural distortions in BiMn$_2$0$_5$ caused by
Bi ions due to the presence of alone pairs of 6s$^2$ electrons ~\cite{6s2}.
The Bi$^{3+}$ ions are nonmagnetic ones but they strongly distort the lattice,
while strongly magnetic Gd$^{3+}$ ions with the ground state ($^8$S$_{7/2}$)
interact weakly with the lattice. Of particular interest is the correlation between
the properties of phase separation domains and electrical polarizations in these
two model crystals. Magnetization, electric polarization, permittivity, and
conductivity studies and also features of high-resolution X-ray diffraction scattering
of RMn$_2$O$_5$ ($R = Gd$, $Bi$) are reported.

A characteristic feature of RMn$_2$O$_5$ is the presence of equal amounts of manganese ions with different valences
Mn$^{3+}$ (containing 3t$_{2g}$, 1e$_g$ electrons in the $3d$ shell) and Mn$^{4+}$ (with 3t$_{2g}$, 0e$_g$ electrons),
which provides conditions for charge ordering. The Mn$^{4+}$ ions have an octahedral oxygen environment
and are arranged in the layers with $z = 0.25c$ and $(1-z)=0.75c$. The Mn$^{3+}$ ions have a local
non-central surrounding in the form of a pentagonal pyramid
and are arranged in the layers with $z = 0.5c$. The R$^{3+}$ ions have the environment
similar to that of the Mn$^{3+}$ ions and
are arranged in the layers $z = 0$ ~\cite{Radaelli2008}.
Charge ordering in RMn$_2$O$_5$ and the e$_g$ electron transfer between Mn$^{3+}$ -Mn$^{4+}$ ion pairs
(double exchange ~\cite{Gannes,Gorkov}) are key factors responsible for polar electric states
of these multiferroics at all temperatures. The low-temperature ferroelectric ordering in RMn$_2$O$_5$
is caused by charge ordering of alternating Mn$^{3+}$ - Mn$^{4+}$ ion pairs with
antiferromagnetically and ferromagnetically oriented spins along the $b$ axis.
Different values of indirect exchange between the antiferromagnetic Mn$^{3+}$ - Mn$^{4+}$ ion pairs
and considerably stronger double exchange between the ferromagnetic pairs of these ions lead to
the exchange striction that breaks the centrosymmetricity of the lattice, thus giving rise to
the ferroelectric polarization along the $b$ axis at $T\leq T_C$ ~\cite{Brink2008,Radaelli2008}.

The double exchange between Mn$^{3+}$-Mn$^{4+}$ ions induces  phase separation
in RMn$_2$O$_5$ which is similar to phase separation in LnAMnO$_3$ (A- Sr, Ba, Ca) manganites
containing Mn$^{3+}$ and Mn$^{4+}$ ions as well. Phase separation exists at all temperatures
and makes the formation of local conductive domains  containing Mn$^{3+}$ - Mn$^{4+}$ ion
pairs with ferromagnetically oriented spins energetically favorable. The phase separation domains
are located in a dielectric antiferromagnetic (paramagnetic) matrix of the original crystal.
The size and geometry of the local phase separation domains are determined by a dynamic balance
of interactions which leads to the attraction of charge carriers (double exchange, the Jahn-Teller effect)
and Coulomb repulsion ~\cite{Gorkov,Kagan}. The effect of phase separation domains on the RMn$_2$O$_5$
properties was studied from 5 K up to the temperatures above room temperature.
At $T < T_C$ the phase  separation domains manifested themselves as $1D$ superlattices
with alternating ferromagnetic layers containing different numbers of charge carriers.
It was supposed that these superlattices were multiferroic domain walls (with their own polarizations)
located between ferroelectric domains of the basic crystal matrix. A set of ferromagnetic
resonances from individual superlattice layers was detected for a number of multiferroics
RMn$_2$O$_5$ ($R = Eu$, $Gd$, $Er$, $Tb$, $Bi$) ~\cite{JETPLett2012,JPCM2012,JPCS2014}.
In the paramagnetic and paraelectric temperature range, up to temperatures above room temperature,
the polar nature of local phase separation domains was found to be responsible for dielectric
properties and the features of the high-resolution X-ray diffraction
spectra of RMn$_2$O$_5$ ($R = Eu$ and $Gd$) single crystals ~\cite{PRB2009,JPCM2011,JETPLett2014}.
The 2D superstructures of alternating layers of the initial RMn$_2$O$_5$ crystals and phase separation
domains were revealed at sufficiently high temperatures. Most distinctly these superstructures
(perpendicular to the c axis) manifested themselves in the X-ray studies of EuMn$_2$O$_5$
and Eu$_{0.8}$Ce$_{0.2}$Mn$_2$O$_5$ at room temperature. The superstructure layer thicknesses
were 900~\AA~ and 700~\AA, respectively ~\cite{PRB2009,JPCM2011}.

\section{Experimental}
\label{experimental}

Single crystals of RMn$_2$O$_5$ were grown by the spontaneous crystallization technique
described in ~\cite{1988,1992}.
The as-grown single crystals were in the form of  2 -- 3 mm thick plates with areas of 3 -- 5 mm$^2$.
To measure the polarization, capacitors with a thickness of 0.3 -- 0.6 mm and area of 3 -- 4 mm$^2$ were used.
The electric polarization was measured by two methods, i.e., the thermo-stimulated pyrocurrent method and
the so-called positive-up negative-down (PUND) method of hysteresis loops measuring
\cite{Scott1988,Fukunaga2008,Feng2010}. In the first case the polarization was measured by a Keithly 6514 electrometer
and the measurements were produced during sample heating with a constant temperature variation rate after the preliminary
sample cooling in a polarizing electric field which was switched off at the lowest temperature. In the second case
we employed the version of the PUND method presented in \cite{Feng2010}
which was adapted to our measurements (see Fig. 1 in \cite{JETPLett2016}).
If the sample had a relatively high conductivity
(which is important for RMn$_2$O$_5$ containing restricted polar phase separation domains with local conductivities),
the shape of the polarization -- electric
field  (P-E) hysteresis loop was distorted and did not give correct information on the intrinsic polarization P.
In the PUND method only the hysteresis of
P can be extracted by  applying a series of voltage pulses to the sample.
During successive positive P1-P2 and negative N1-N2 pulses, independent curves (P1-P2 and N1-N2)
of effective polarization P changes are registered. The PUND method is based on the difference
between polarization and conductivity responses to variations in the field E. The time intervals between
P1-P2 and N1-N2 pulses should be chosen such that the intrinsic polarization P is still unrelaxed, while
the conductivity relaxation is complete. In the conventional volume ferroelectrics with the domain structure
those time intervals may be up to seconds. In our case the intrinsic P response was determined by the restricted
polar phase separation domains which rather rapidly restored after the field E was switched off.
The reason for this will be discussed below.
%
\begin{figure}[htb]
  \includegraphics[width=0.45\textwidth, angle=0]{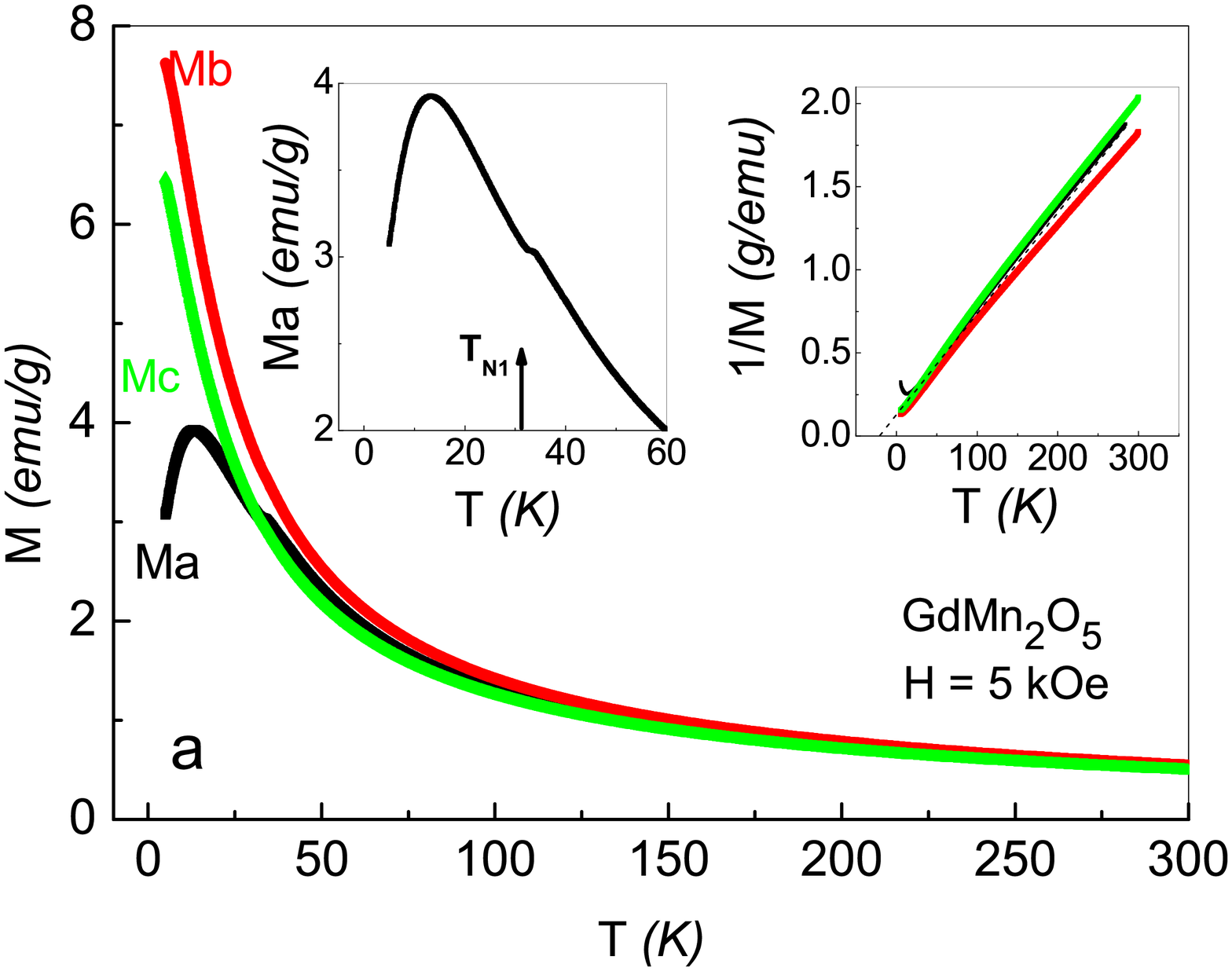}
  \includegraphics[width=0.45\textwidth, angle=0]{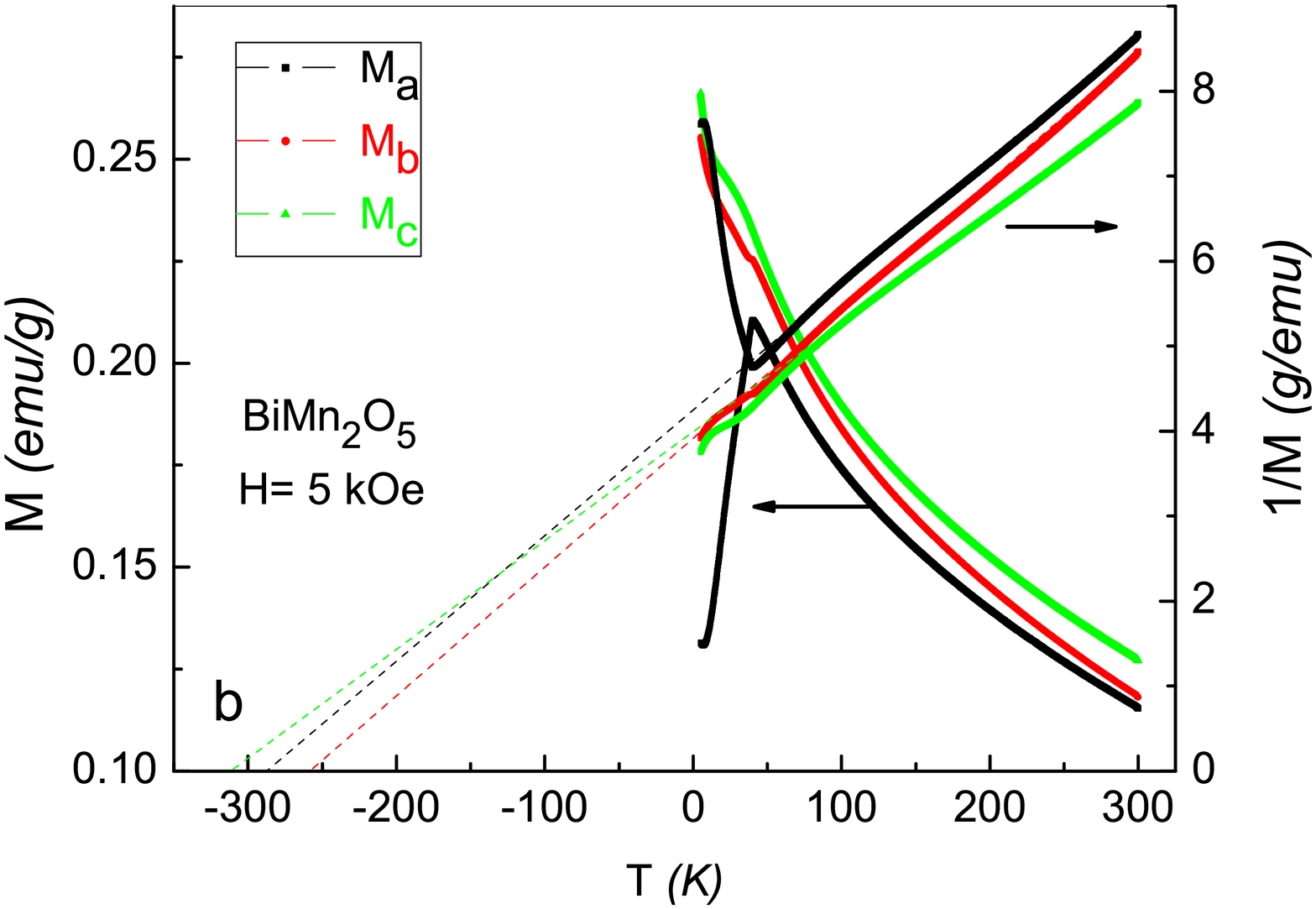}
  \caption{Temperature dependences of magnetization along all crystal axes for GdMn$_2$O$_5$ (a) and BiMn$_2$O$_5$ (b).}
\label{fig1}
\end{figure}
%
\noindent As a result, the time intervals between the P1-P2 and N1-N2 pulses were chosen so that the responses
to the P1 and N1 pulses were irreversible (due to the intrinsic polarization contribution) and the conductivity responses
to the P2 and N2 pulses were closed and reversible (see Fig.~\ref{fig1} in \cite{JETPLett2016}). This could be achieved if
the time intervals between the P1-P2 and N1-N2 pulses did not exceed 0.8 ms.
The P1 and N1 curves characterized the total
polarization and conductivity contributions.
To obtain the actual P-E loop, we subtracted the P2 and N2 curves from the P1-N1 curves.
The durations of pulses in the P1-P2 and N1-N2 series were 2 ms and 4.5 ms, the intervals between the pulse series
were 4 ms and 9 ms for GdMn$_2$O$_5$ and BiMn$_2$O$_5$, respectively.
The permittivity and conductivity were measured by a Good Will LCR-819
impedance meter in the frequency range 0.5 -- 50 kHz at 5 -- 350 K. Magnetic properties were investigated by
PPMS (Quantum Design). The intensity distributions of Bragg reflections
were studied with a high-sensitivity 3-crystal X-ray diffractometer.

\section{Experimental results and analysis}
\label{results}
\subsection{Magnetic properties and low-temperature ferroelectric state}
\label{magnetic}

Complicated magnetic structures with wave vector $q = (1/2+\delta x, 0, 1/4+\delta z)$ and
a number of low-temperature magnetic phase transitions
at $T\leq T_N$ at which incommensuration parameters $\delta x$ and $\delta z$ change in a step-like fashion
are typically observed in RMn$_2$O$_5$ with different R ions ~\cite{Radaelli2008}. As temperature decreases,
an incommensurate magnetic phase is appeared near T$_N$.
At a lower temperature this phase is transformed into a commensurate structure. As temperature further decreases,
one more transition into the incommensurate phase is possible. Ferroelectric ordering in RMn$_2$O$_5$
is typically occurs in the intermediate commensurate phase.

Magnetic structures of GdMn$_2$O$_5$ and BiMn$_2$O$_5$ differ from those commonly observed in RMn$_2$O$_5$.
In GdMn$_2$O$_5$, a commensurate collinear antiferromagnetic structure with wave vector $q = (1/2, 0, 0)$ occurs
in the temperature range 0 -- 30 K \cite{LeePRL}.
At these temperatures there is a ferroelectric ordering with $T_C\simeq 30$ K. In the temperature range
35 K$\geq T > 30$ K there is an incommensurate phase. In BiMn$_2$O$_5$ the commensurate noncollinear
antiferromagnetic structure with $q = (1/2, 0, 1.2)$ is observed at
T$\leq T_N\simeq 40$ K \cite{Radaelli2008,Munoz}.
%
\begin{figure}[htb]
  \includegraphics[width=0.45\textwidth, angle=0]{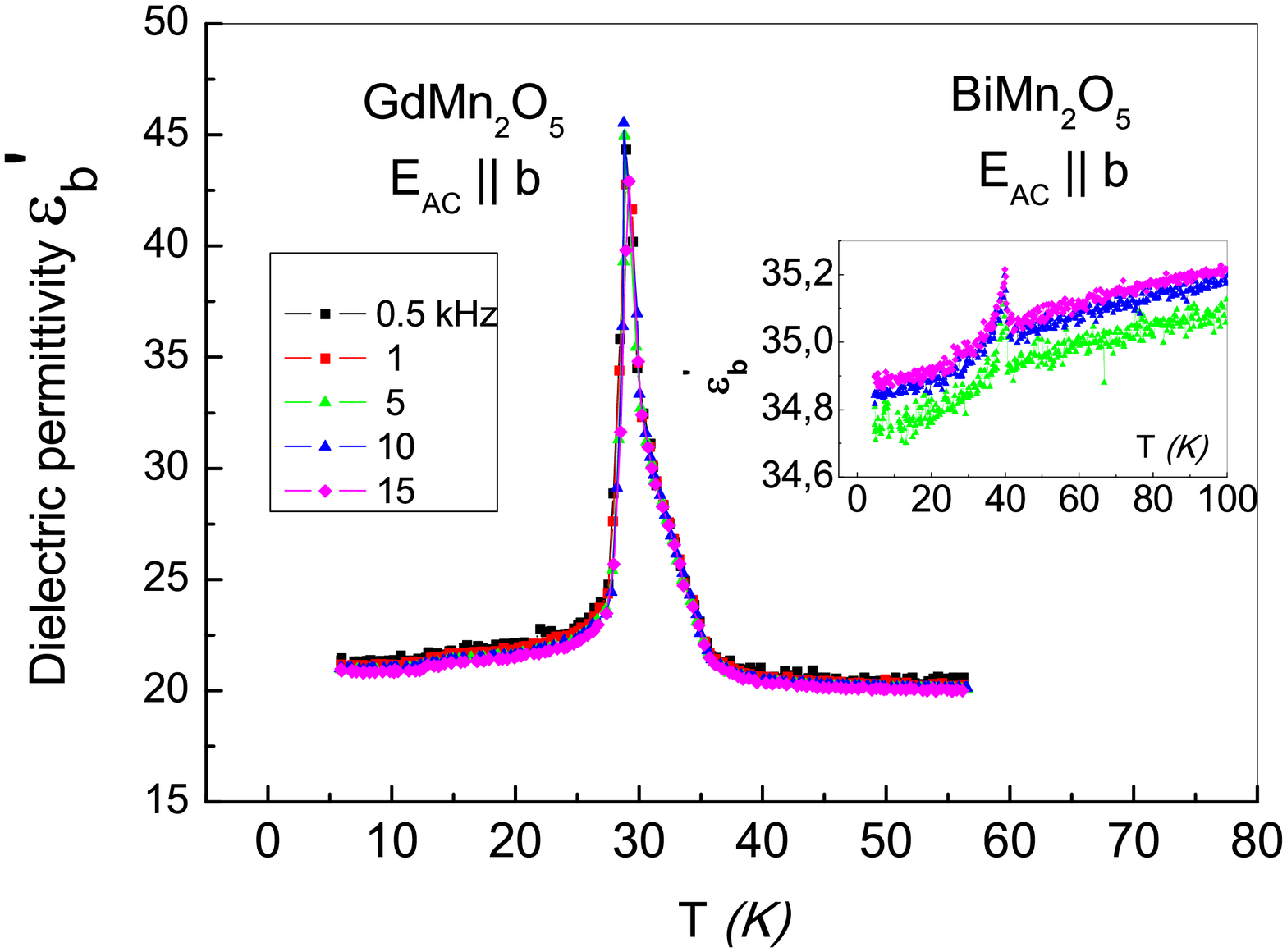}
  \caption{Dispersion-free anomalies of real permittivity $\varepsilon^{\prime}$ along the $b$ axis at $T\simeq T_C$ for GdMn$_2$O$_5$ and BiMn$_2$O$_5$ (in the inset).}
\label{fig2}
\end{figure}
%
\noindent Fig.~\ref{fig1}  shows temperature dependence of magnetization in magnetic field $H = 5$ kOe along different
axes of the GdMn$_2$O$_5$ (Fig.~\ref{fig1}a)
and BiMn$_2$O$_5$ (Fig.~\ref{fig1}b) crystals studied. It can be seen that the magnetic moment
in GdMn$_2$O$_5$ is much higher due to the contribution of Gd$^{3+}$ ions with the $S = 7/2$ spin which
have their own ordering at $T\simeq 13$ K.
On the background of magnetization of Gd ions, the antiferromagnetism of manganese ions along the a axis
(with $T_N = 35$ K) manifests itself only slightly (see the left inset in Fig.~\ref{fig1}a).
The right inset in Fig.~\ref{fig1}a shows that the Curie-Weiss temperature for GdMn$_2$O$_5$ nearly coincides
with T$_N$. This means that there is a distinct phase transition near T$_N$ in GdMn$_2$O$_5$
(the magnetic state is not frustrated).
A quite different magnetic state is observed in BiMn$_2$O$_5$ (Fig.~\ref{fig1}b).
The antiferromagnetism along the $c$ axis manifests
itself only slightly, while it is pronounced along the $a$ axis. Unlike in GdMn$_2$O$_5$,
but similar to RMn$_2$O$_5$ with other R ions,
the Curie-Weiss temperature exceeds $T_N = 40$ K in 6-7 times.
This means that the magnetic state in BiMn$_2$O$_5$ is highly frustrated,
and spin correlations in local domains are observed at $T\gg T_N$.
Specific magnetic and structural properties of BiMn$_2$O$_5$ in the RMn$_2$O$_5$ family
are caused by the presence of electron alone pairs in Bi$^{3+}$ ions.

High-resolution neutron powder diffraction studies \cite{Munoz} show that BiMn$_2$O$_5$ is characterized at room temperature
by the orthorhombic sp. gr. {\it Pbam}, similar to other RMn$_2$O$_5$. However, Mn$^{4+}$-O6 octahedra, Mn$^{3+}$-O5 pyramids,
and Bi$^{3+}$-O units are more deformed as compared with those in other RMn$_2$O$_5$. As a result, the Mn-O distances
and magnetic exchange bonds differ as well.

\begin{figure}[htb]
  \includegraphics[width=0.45\textwidth, angle=0]{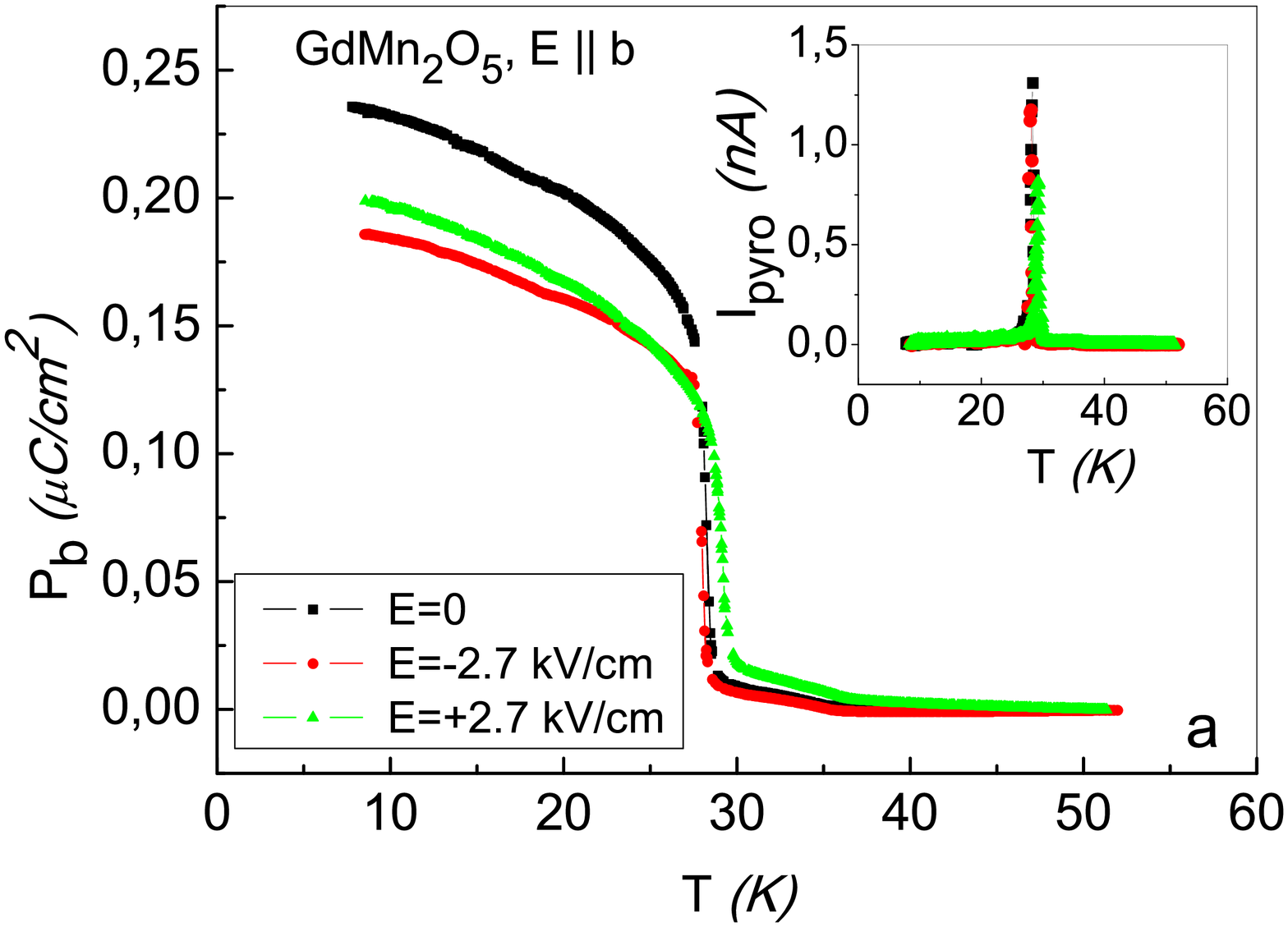}
  \includegraphics[width=0.45\textwidth, angle=0]{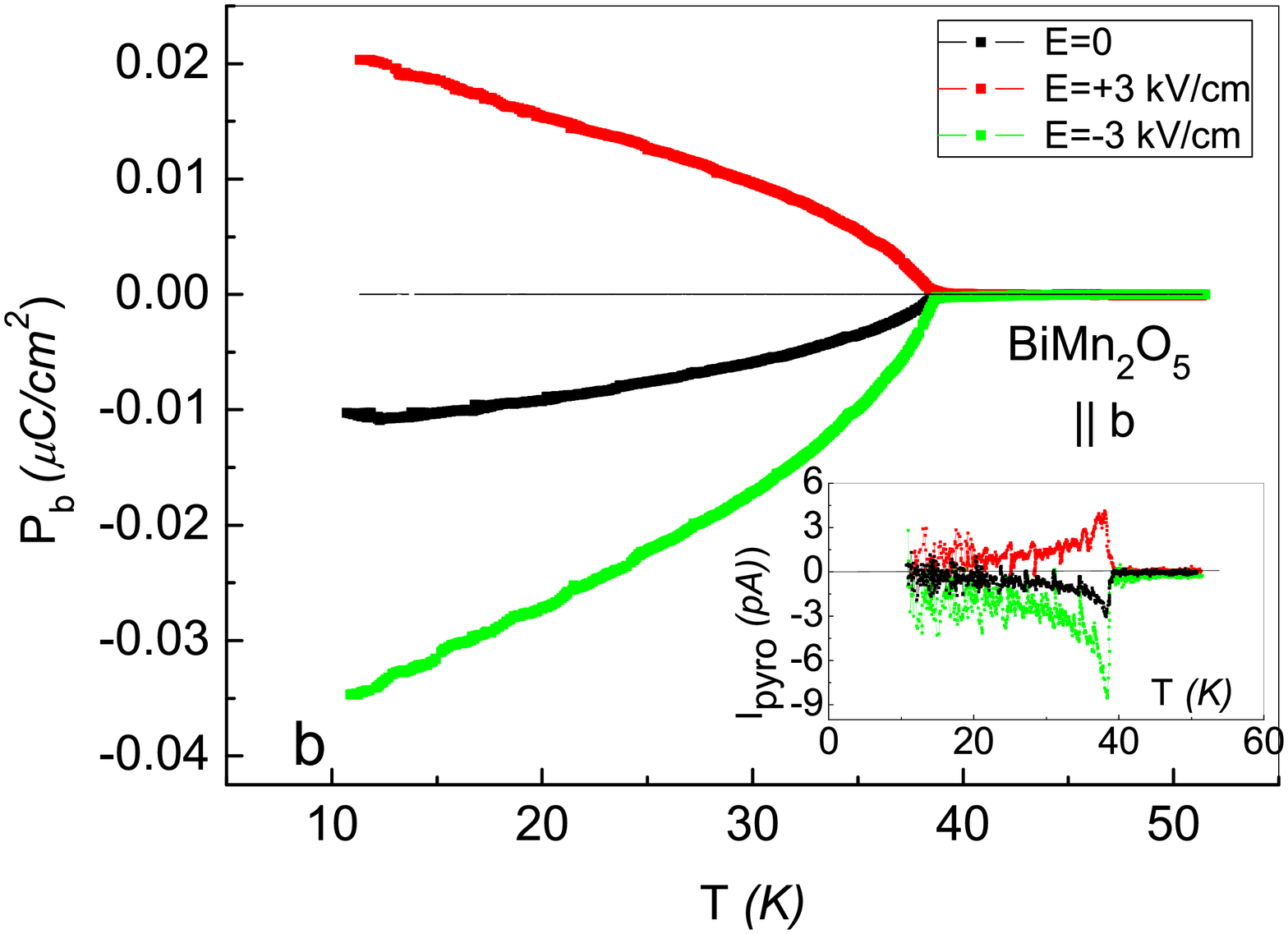}
  \caption{Temperature dependences of polarizations P$_b$ and pyrocurrents along the $b$ axis at $E = 0$ and $E\neq 0$ for
   GdMn$_2$O$_5$ (a) and BiMn$_2$O$_5$ (b).}
\label{fig3}
\end{figure}

As will be shown below, differences in magnetic and structural states of GdMn$_2$O$_5$ and BiMn$_2$O$_5$ lead
to different electric polarizations at all temperatures. Let us consider at first the low-temperature polar states.
Fig.~\ref{fig2} shows dispersion-free anomalies of real permittivity $\varepsilon^{\prime}$ along the b axis
which are characteristic of the low-temperature ferroelectric phase transition. As one can see, the value
of $\varepsilon^{\prime}$ at the maximum in the vicinity of T$_C$ in GdMn$_2$O$_5$ considerably exceeds the maximum in
BiMn$_2$O$_5$. Fig.~\ref{fig3} shows electric polarizations
along the $b$ axis (P$_b$) in GdMn$_2$O$_5$ (Fig.\ref{fig3}a) and BiMn$_2$O$_5$ (Fig.\ref{fig3}b).
The polarizations are measured by the pyrocurrent method on the assumption that P$_b = 0$ at T$\geq T_C$.
The maximum polarization P$_b$ (in our case $P_b\simeq 0.26$ $\mu$C/cm$^2$) for the RMn$_2$O$_5$ family
is observed in GdMn$_2$O$_5$ ~\cite{LeePRL}.
In BiMn$_2$O$_5$, polarization P$_b$ is an order of magnitude lower and is closer to the values
typically observed in other RMn$_2$O$_5$ \cite{Noda2008}.
%
%
\noindent  Note that a fully polarized state is often observed in structurally perfect (according to X-ray diffraction)
RMn$_2$O$_5$ single crystals with different R ions in the absence of an applied external electric field E~\cite{Noda2}.
This indicates that this fully polarized P$_b$ polarization
is formed in the internal field. We suppose that this is a staggered-like field caused by charge ordering
of Mn$^{3+}$ - Mn$^{4+}$ ion pairs along the $b$ axis,
and the polarization has an exchange-striction nature. This situation is observed for both crystals we studied.
In GdMn$_2$O$_5$ (Fig.~\ref{fig3}a), the maximum polarization is detected in $E = 0$. A uniform external
field $\pm E\parallel b$ which is much weaker than the internal staggered-like field,
only slightly reduces polarization P$_b$ and cannot reverse its orientation (Fig.~\ref{fig3}a).
The application of a strong field $E\parallel b$
(comparable to the internal field) enhances the transfer of e$_g$ valence electrons between the
Mn$^{3+}$ -Mn$^{4+}$ ion pairs along the $b$ axis,
which leads to the electric breakdown of the crystal. Therefore, measurements of the polarization hysteresis loops
P$_b$ (which are the response to an applied external field E) is not effective when the striction
contribution to the GdMn$_2$O$_5$ polarization at $T <T_C$ is measured.
%
\begin{figure}[!htb]
  \includegraphics[width=0.45\textwidth, angle=0]{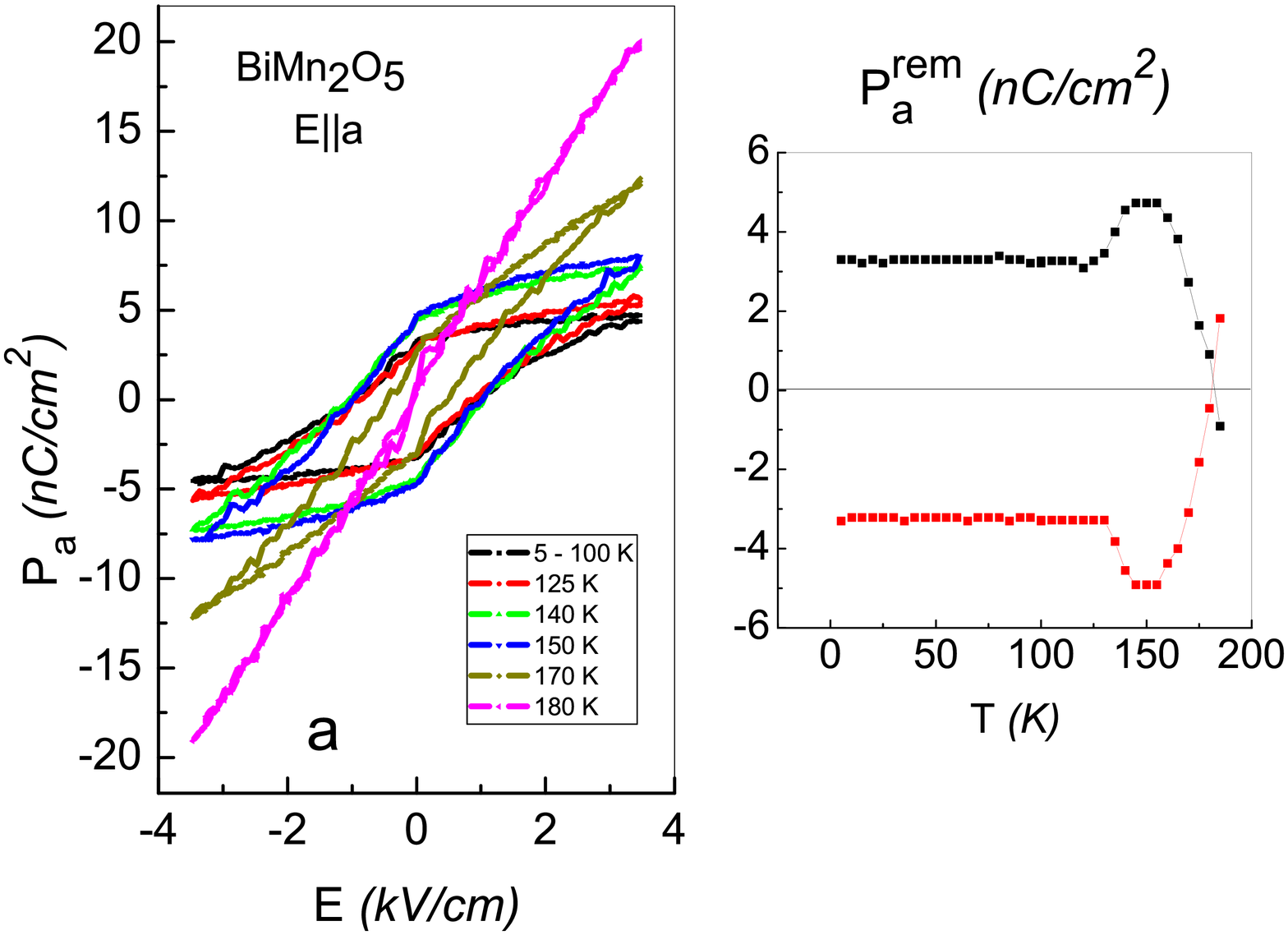}
  \includegraphics[width=0.45\textwidth, angle=0]{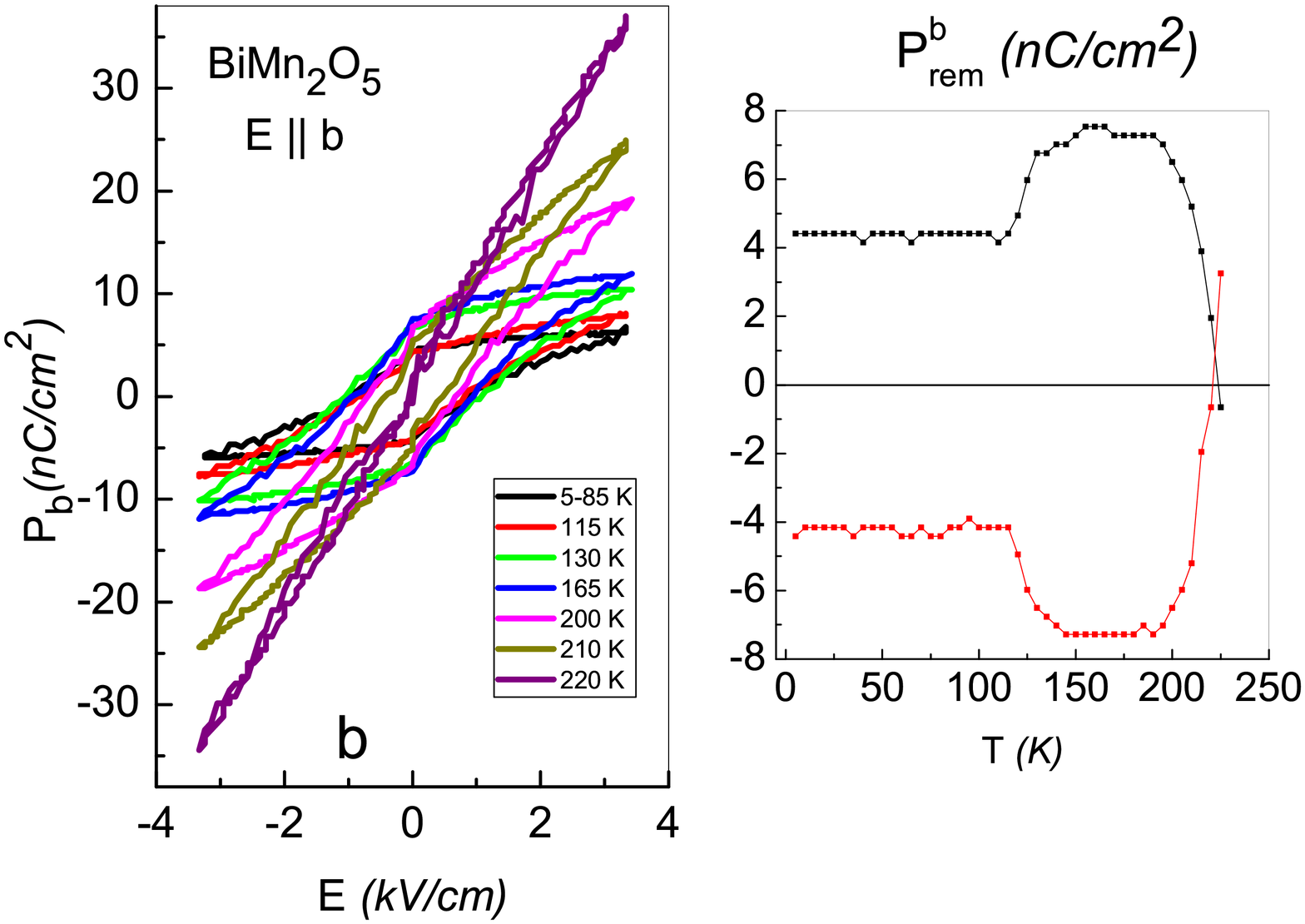}
  \includegraphics[width=0.45\textwidth, angle=0]{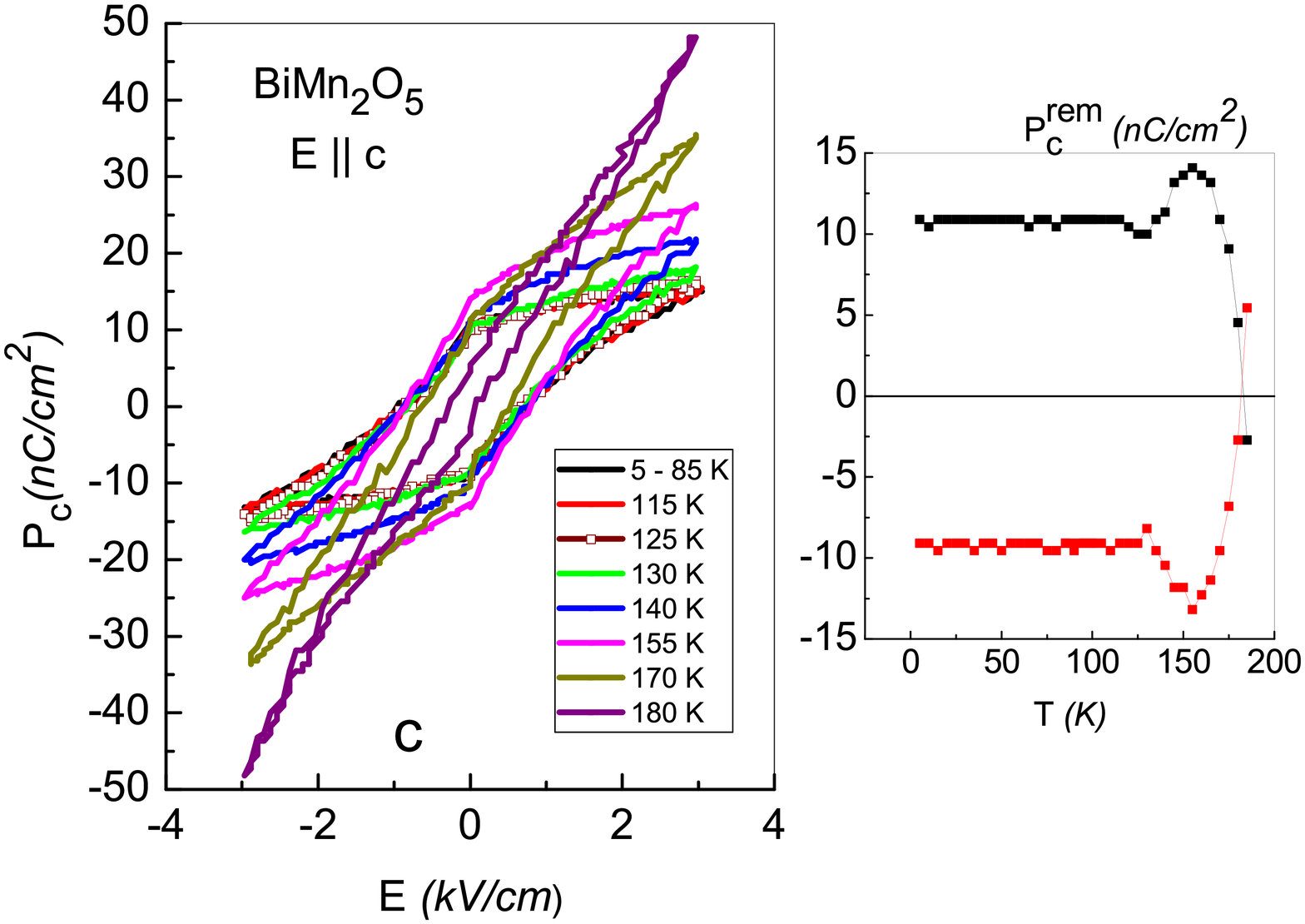}
  \caption{Polarization hysteresis loops for BiMn$_2$O$_5$ (left panels) at different temperatures at E$\parallel a$ (a),
   E$\parallel b$ (b), and E$\parallel c$ (c). Temperature dependences of remanent polarization are shown in right panels.}
\label{fig4}
\end{figure}
%
\noindent In BiMn$_2$O$_5$ (which are also structurally perfect single crystals), Bi ions distorts the uniformityof the internal
staggered-like field, thus reducing its value and, hence, the exchange-striction polarization P$_b$.
In this case application of field $\pm E\parallel b$ changes the internal field to a greater extent and can even reverse the BiMn$_2$O$_5$
polarization (see Fig.~\ref{fig3}b).
As can be seen, the application of E in the same direction in which the polarization was observed at $E = 0$ enhances
the polarization by the value close to that detected in E of the opposite orientation (Fig.~\ref{fig3}b).
There is good reason to believe that the internal
exchange-striction polarization is the polarization observed at $E = 0$. There are two additional
contributions to the polarization of BiMn$_2$O$_5$ in the case of application of field $E\parallel b$.
There is the contribution originating from a weak change in the internal
staggered-like-field and that due to the response of phase separation domains to the applied field E.
Note that the polarization measured
by the pyrocurrent method includes the contribution of conductivities of phase separation domains.

\subsection{Polarization hysteresis loops for GdMn$_2$O$_5$ and BiMn$_2$O$_5$}
\label{loops}

As noted above, it was shown in \cite{JETPLett2016} that electric polarization hysteresis loops
in GdMn$_2$O$_5$ were observed in a wide temperature
interval (from the lowest up to room temperature). We put forward the hypothesis that this polarization
was due to the frozen superparaelectric state of restricted polar phase separation domains formed
in the initial crystal matrix. Phase separation must exist in RMn$_2$O$_5$ with any R ions
at all temperatures, but the phase separation domain states depend on the R ion type and temperature.
Thus, two electric polarizations having different natures (the long-range ferroelectric order
with polarization P$_b$ caused by exchange striction and the polarization due to
polar phase separation domains) must coexist in RMn$_2$0$_5$ along the b axis at $T\leq T_C$.
The previous subsection was concerned with the polarizations P$_b$ at $T\leq T_C$ measured
by the pyrocurrent method at $E = 0$ in GdMn$_2$O$_5$ (Fig.\ref{fig3}a) and BiMn$_2$O$_5$ (Fig.~\ref{fig3}b).
This polarization gave no contribution to the hysteresis loops which were responses to the applied field E.
On the contrary, the polarization caused by phase separation domains could be determined only by measuring
the hysteresis loops by the PUND method that excluded the contribution of conductivity. By using these two methods,
we could separate the contributions of the polarizations considered above at low temperatures.

Let us consider in more detail the hysteresis loops for BiMn$_2$O$_5$ in a wide temperature interval and
compare them with the loops for GdMb$_2$O$_5$. Fig.~\ref{fig4} shows the P-E hysteresis loops of BiMn$_2$O$_5$
in E oriented along the $a$, $b$, and $c$ axes (left panels in Figs.~\ref{fig4}a,~\ref{fig4}b,
and ~\ref{fig4}c, respectively). The right panels in these figures demonstrate temperature
dependences of the remanent polarization ($P^{rem}$).
%
\begin{figure}[!htb]
  \includegraphics[width=0.45\textwidth, angle=0]{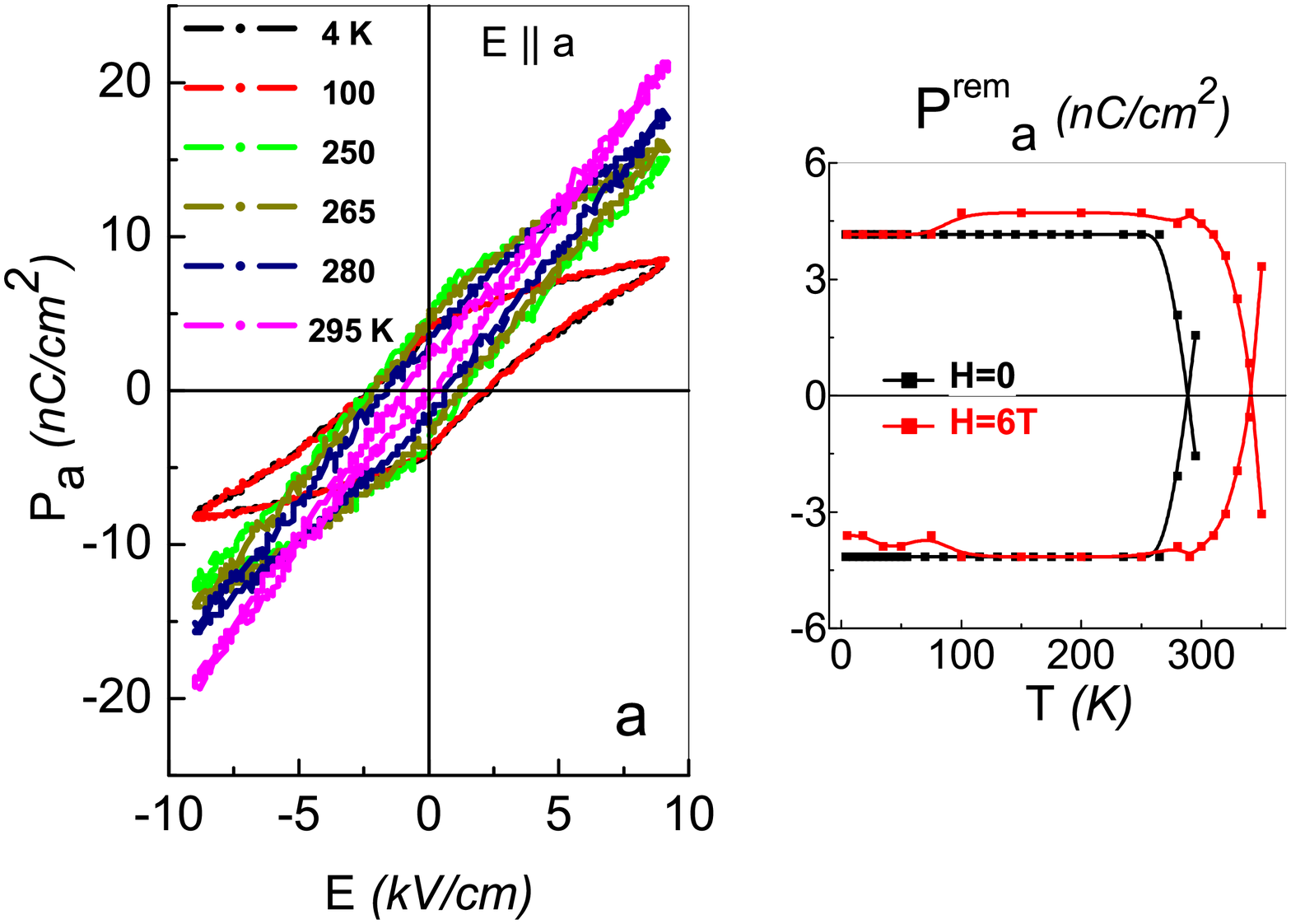}
  \includegraphics[width=0.45\textwidth, angle=0]{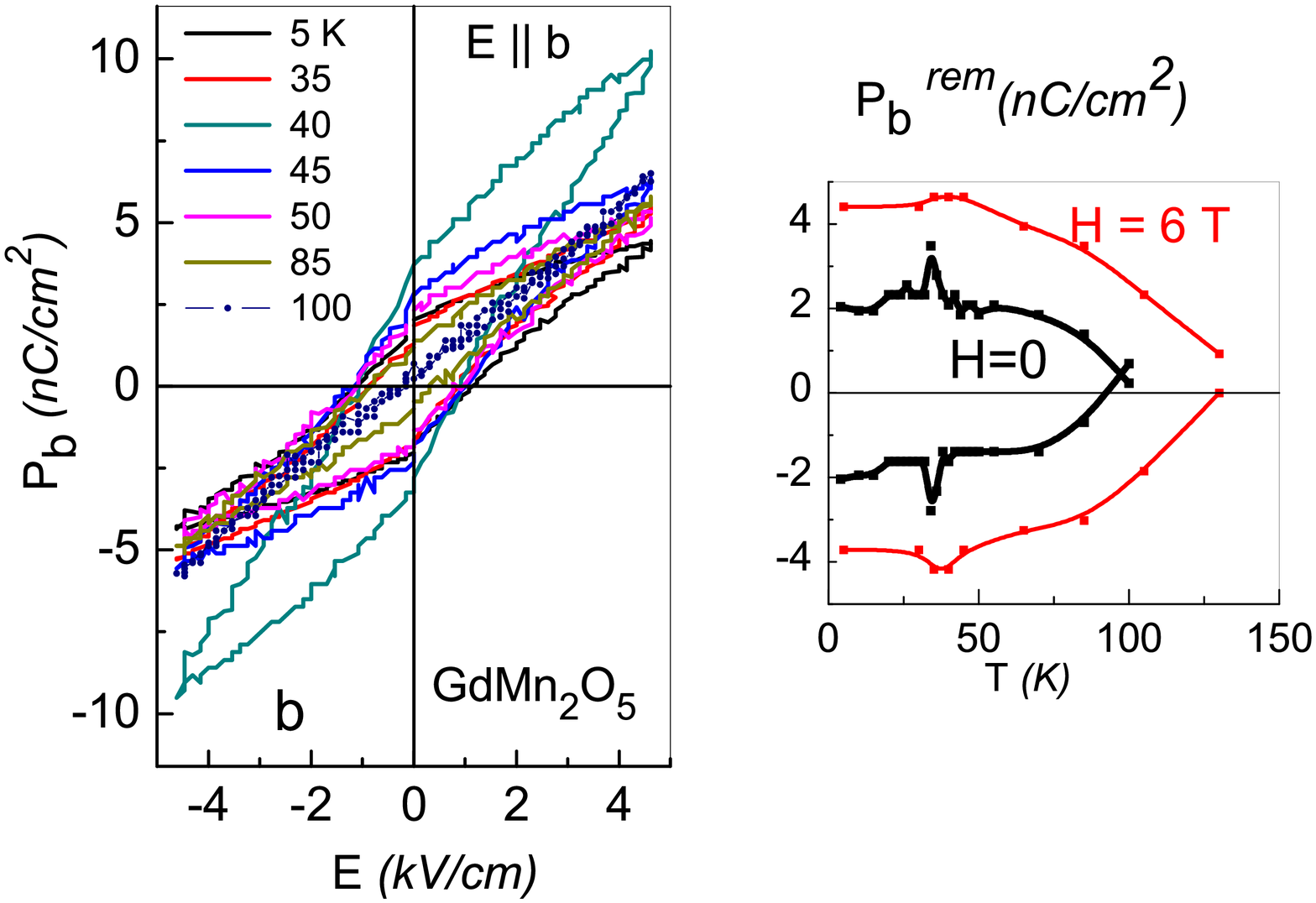}
  \includegraphics[width=0.45\textwidth, angle=0]{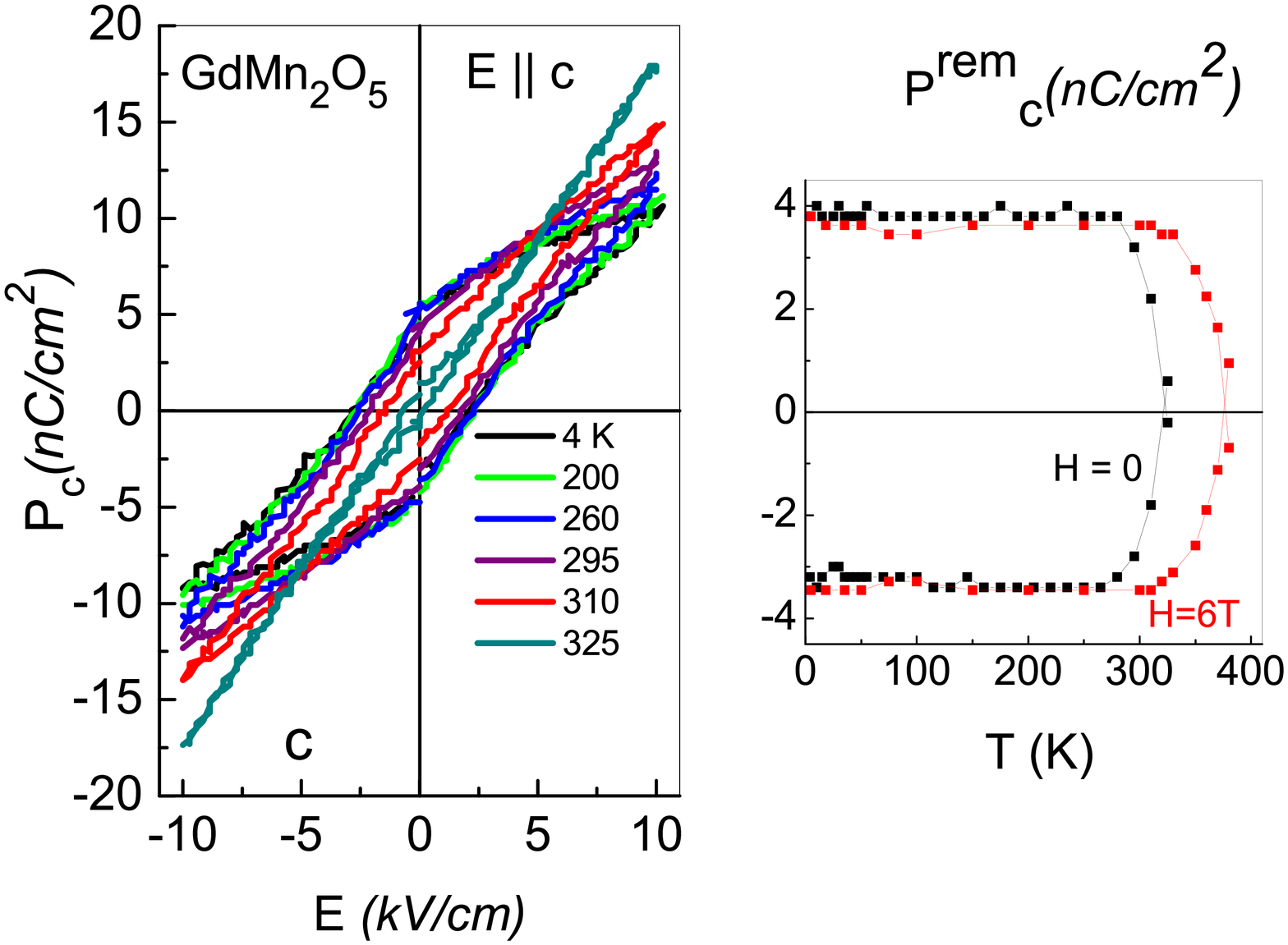}
  \caption{Polarization hysteresis loops for GdMn$_2$O$_5$ (left panels) at different temperatures at E$\parallel a$ (a),
   E$\parallel b$  (b), and E$\parallel c$ (c). Temperature dependences of remanent polarization are shown
   in right panels.}
\label{fig5}
\end{figure}
%
\noindent  Figs. ~\ref{fig5}a, ~\ref{fig5}b, and ~\ref{fig5}c present the P-E hysteresis loops and Prem for GdMn$_2$O$_5$.
As one can see, the hysteresis loops are observed for both crystals along all three axes from 5 K up to some
temperatures T$^*$ which depend on the axis direction.
The maximum $P^{rem}$ polarizations are observed along the $c$ axis from 5 K up to $T^*\simeq 180$ K in BiMn$_2$O$_5$ and
up to $T^*\simeq 325$ K for GdMn$_2$O$_5$. The minimum $P^{rem}$ along the $b$ axis which exists up
to $T^*\simeq 100$ K is observed for GdMn$_2$O$_5$,
while a twice higher $P^{rem}$ along the $b$ axis for BiMn$_2$O$_5$ manifests itself up to $T^*\simeq 220$ K.
The $P^{rem}$ along the $a$ axis are observed for GdMn$_2$O$_5$ and BiMn$_2$O$_5$ up to $T^*\simeq 295$ K
and $T^*\simeq 180$ K, respectively.
Thus, $P^{rem}$ and hysteresis loops for BiMn$_2$O$_5$ and GdMn$_2$O$_5$ which demonstrate a strong anisotropy
are revealed in the paramagnetic phase. It is important to note that the polarizations of both crystals
occurred along all crystal axes and disappeared at different temperatures along these axes.
This fact means that the polarizations measured in the hysteresis loops were not caused by
the ferroelectric phase transition in the homogeneous
ferroelectric crystal states. As noted above, we assumed that these polarizations
were induced by the restricted polar phase separation domains. It turned out that these polarizations
were significantly lower than the exchange-striction ferroelectric low-temperature polarization
along the $b$ axis at $T\leq T_C$.

The remanent polarization $P^{rem}$ along the $b$ axis (Fig.~\ref{fig5}b) in GdMn$_2$O$_5$ is two orders
of magnitude lower than the exchange-striction contribution
of the basic crystal matrix polarization $P_b$ at $T <30$ K measured by the pyrocurrent method
in the same sample at $E = 0$ (Fig.~\ref{fig3}a).
The contribution of $P_b$ into the hysteresis loops manifests itself only in the form
of maxima on the background of $P^{rem}_b$ near $T_C$ when the internal staggered-like-field which
gives rise to polarization $P_b$ begins to decrease and disappears and the $P_b$ fluctuations
increase (see Fig.~\ref{fig5}b). There is no exchange-striction polarization of the main matrix
in the direction of the $a$ and $c$ axes at $T\leq T_C$, and therefore contributions to the hysteresis
loops come only from phase separation domains at all temperatures.
The fact that $P^{rem}$ is independent of temperature along all axes of GdMn$_2$O$_5$ in a wide temperature range
(for the $b$ axis this relates to the background value minus the maxima near $T_C$) indicates that the
contribution into the polarization from phase separation domains in GdMn$_2$O$_5$ is anisotropic
but temperature-independent. At $T^*$ the remanent polarizations
rather abruptly disappear along different crystal axes. Of interest is the fact that,
unlike in GdMn$_2$O$_5$, the $P^{rem}$ values along all axes of the BiMn$_2$O$_5$ crystal
abruptly increase in the vicinity of $T\simeq 125$ K and then rapidly reduce to zero near
the $T^*$. Such $P^{rem}$ jumps near 125 K indicate that phase separation domain states
in BiMn$_2$O$_5$ should change in a step-like fashion
near this temperature. The absence of such jumps in GdMn$_2$O$_5$ shows that most probably these jumps in BiMn$_2$O$_5$
result from structural lattice distortions caused by the Bi ions.

 The effect of the longitudinal magnetic field H on the hysteresis loops was also studied. As one can see from
the right panels in Fig.~\ref{fig5}, the field $H = 6$ T increases in GdMn$_2$O$_5$ both the remanent polarization
and temperature $T^*$ along the $b$ axis and only $T^*$ along the $a$ and $c$ axes.

Characteristic activation barriers E$_A$ must exist at the interfaces between phase separation domains
and the main matrix of the crystal. Information on E$_A$ along different crystal axes can be obtained
from frequency-temperature dependences of real permittivity $\varepsilon^{\prime}$
and conductivity $\sigma$.
%
\begin{figure}[!htb]
  \includegraphics[width=0.45\textwidth, angle=0]{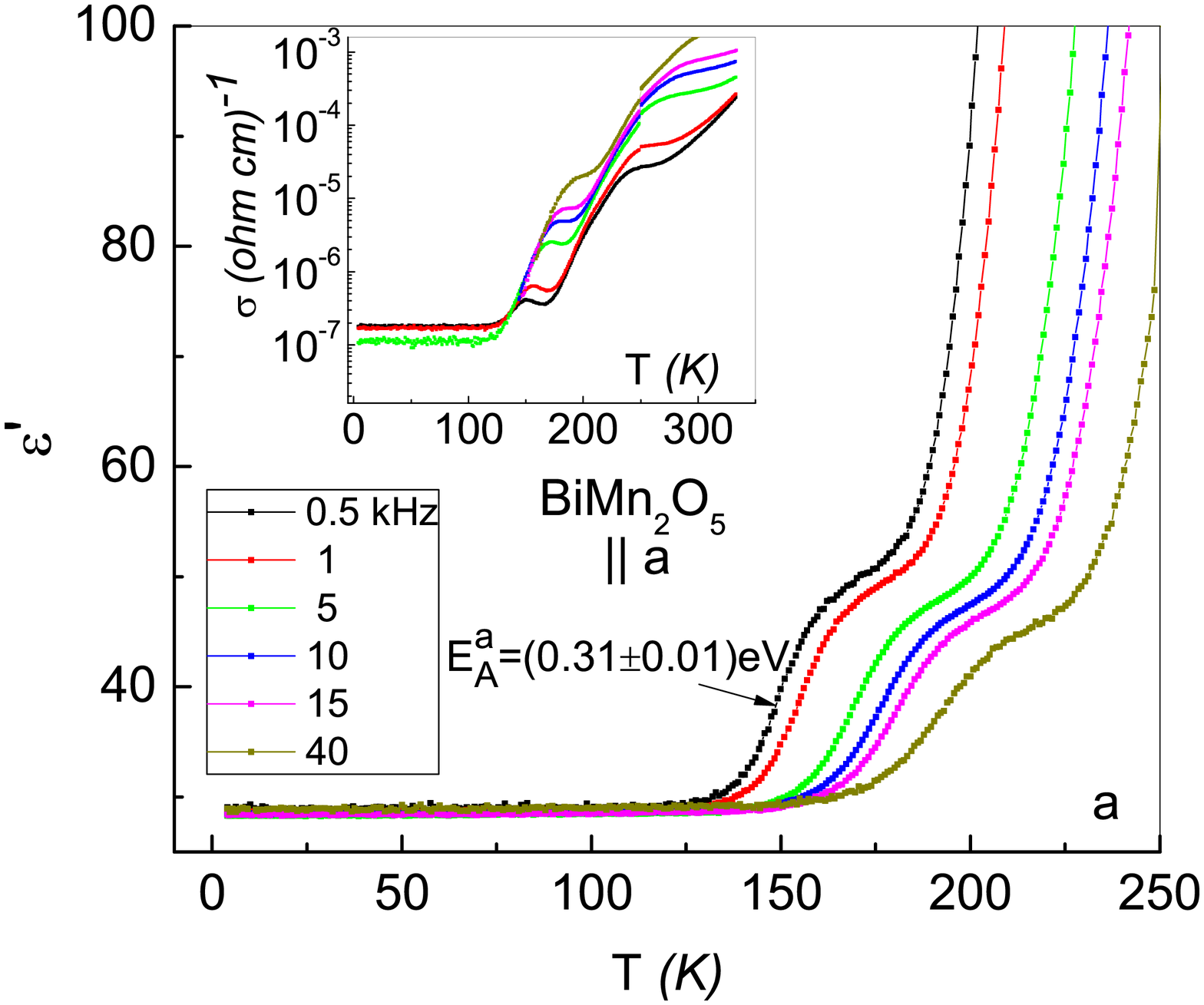}
  \includegraphics[width=0.45\textwidth, angle=0]{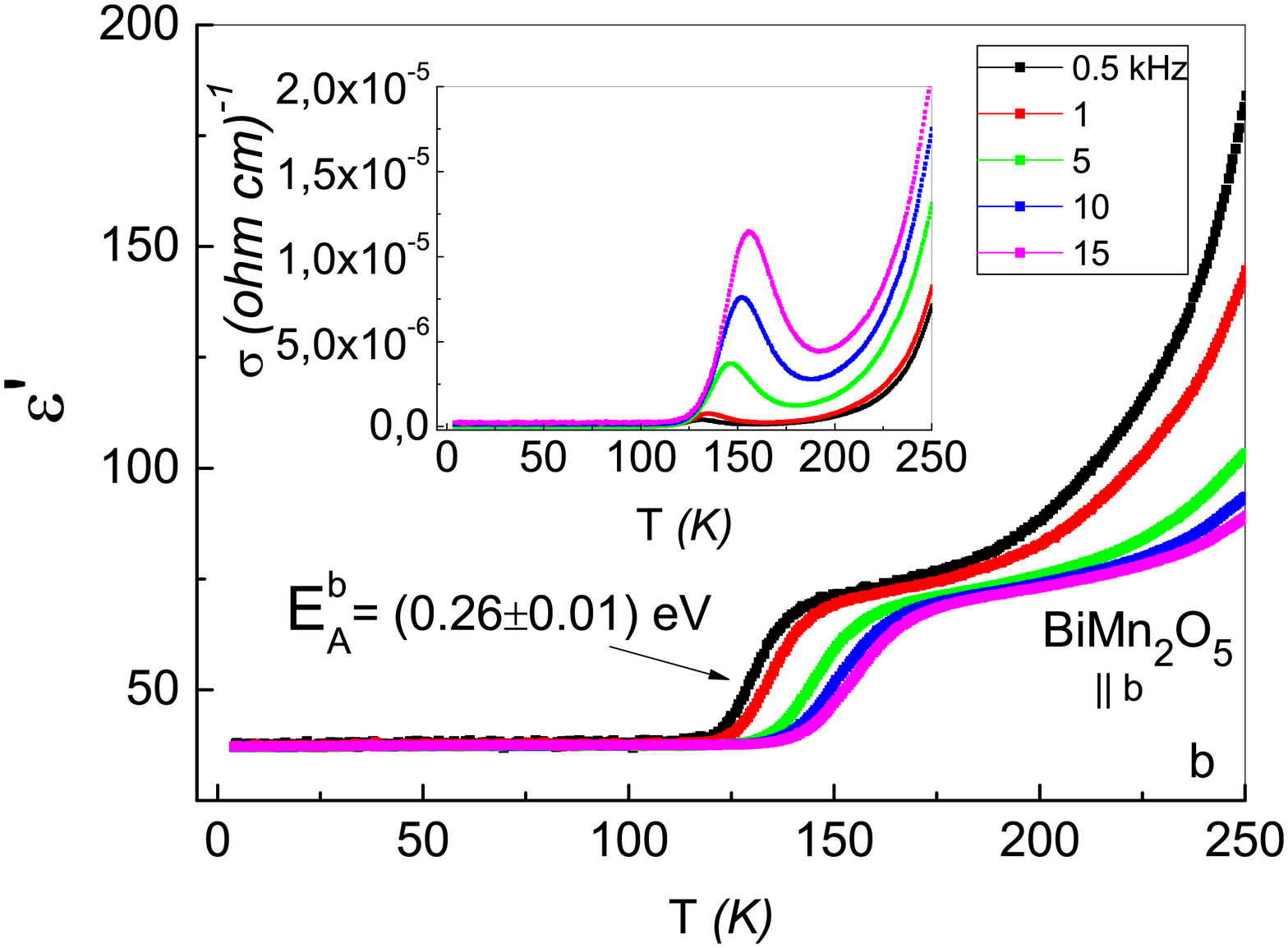}
  \includegraphics[width=0.45\textwidth, angle=0]{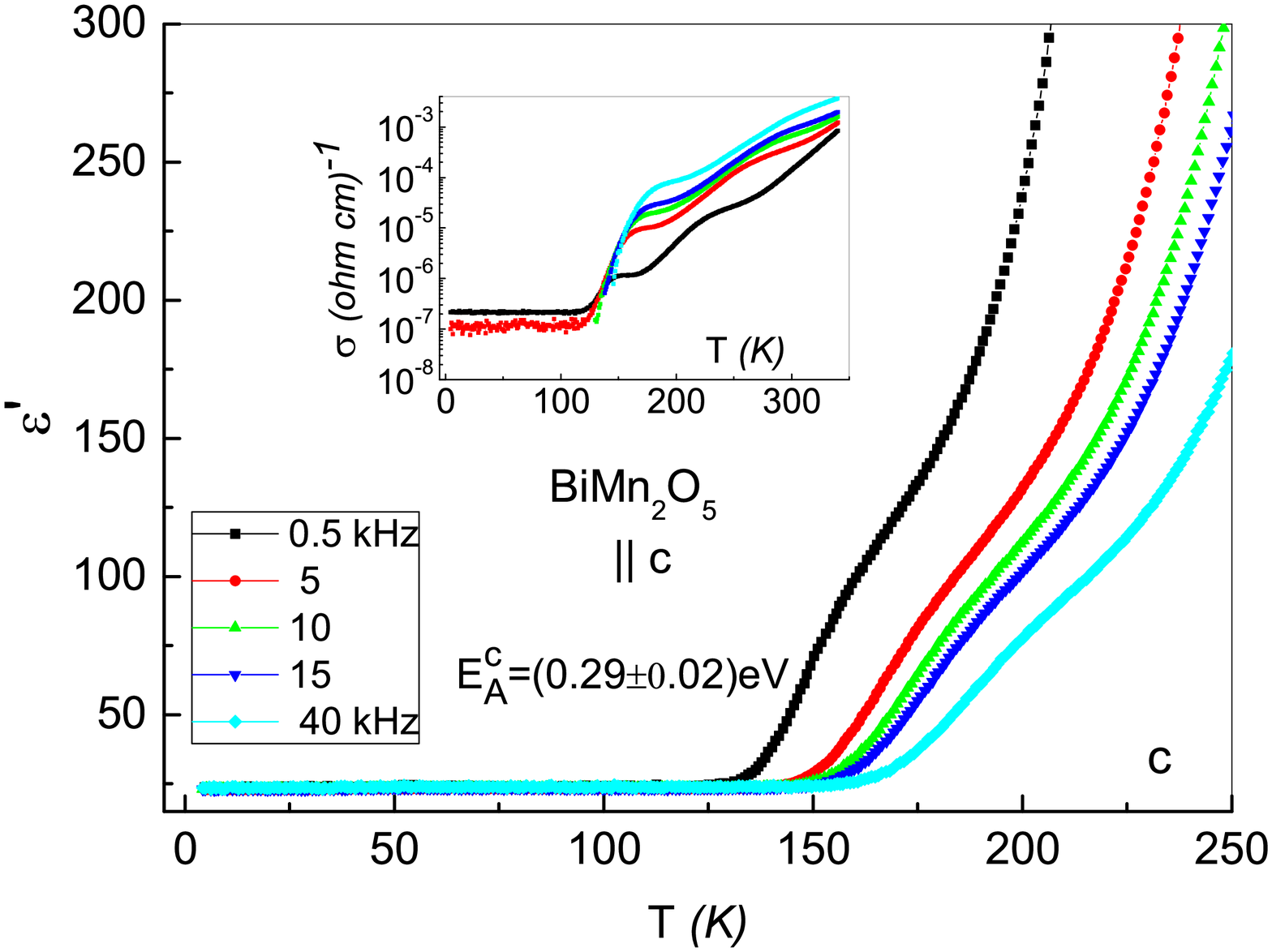}
  \caption{Temperature dependences of the real permittivity $\varepsilon^{\prime}$ and conductivity $\sigma$ (in insets) at different frequencies
  for BiMn$_2$O$_5$ along the $a$ axis (a), $b$ axis (b), and  $c$ axis (c).}
\label{fig6}
\end{figure}
%
\noindent Fig.~\ref{fig6} shows the $\varepsilon^{\prime}$(T) and $\sigma$(T) dependences
for BiMn$_2$O$_5$ at different frequencies along the $a$ axis (Fig.~\ref{fig6}a), $b$ axis (Fig.~\ref{fig6}b),
and $c$ axis (Fig.~\ref{fig6}c). It is evident that the values of $\varepsilon^{\prime}$
and $\sigma$ along all axes of the crystal are nearly independent of temperature at $T\leq 125$ K.
Near $T\simeq 125$ K a frequency-dependent step-like $\varepsilon^{\prime}$ increase to some
constant level (most pronounced along the $a$ and $b$ axes) begins.
As temperature further grows, $\varepsilon^{\prime}$ starts to increase sharply again.
The frequency-dependent growth of $\sigma$ also begins near 125 K,
and two the different temperature intervals of $\sigma$ increase are observed as well. Note that
the temperature independence of $\varepsilon^{\prime}$
and $\sigma$ at $T\leq 125$ K and similar steps in the $\varepsilon^{\prime}$(T) and $\sigma$(T) dependences
near 125 K correlate with the temperature dependences of $P^{rem}$ in BiMn$_2$O$_5$ along different crystal axes
(see Fig.~\ref{fig4}). Indeed, near $T\simeq 125$ K $P^{rem}$ increases
in a step-like fashion along all axes of the crystal. The temperatures at which the second high-temperature
growth in $\varepsilon^{\prime}$ and $\sigma$
start are close to $T^*$ for different crystal axes. The temperatures of $\varepsilon^{\prime}$ jumps
at different frequencies at $T^* > T >125$ K
are well described by the Arrhenius law, which allows one to estimate characteristic activation
barriers $E_A$ at the phase separation domain boundaries
in BiMn$_2$O$_5$. These values are close to $0.3$ eV along different crystal axes (see Fig.~\ref{fig6}).

\begin{figure}[!htb]
  \includegraphics[width=0.45\textwidth, angle=0]{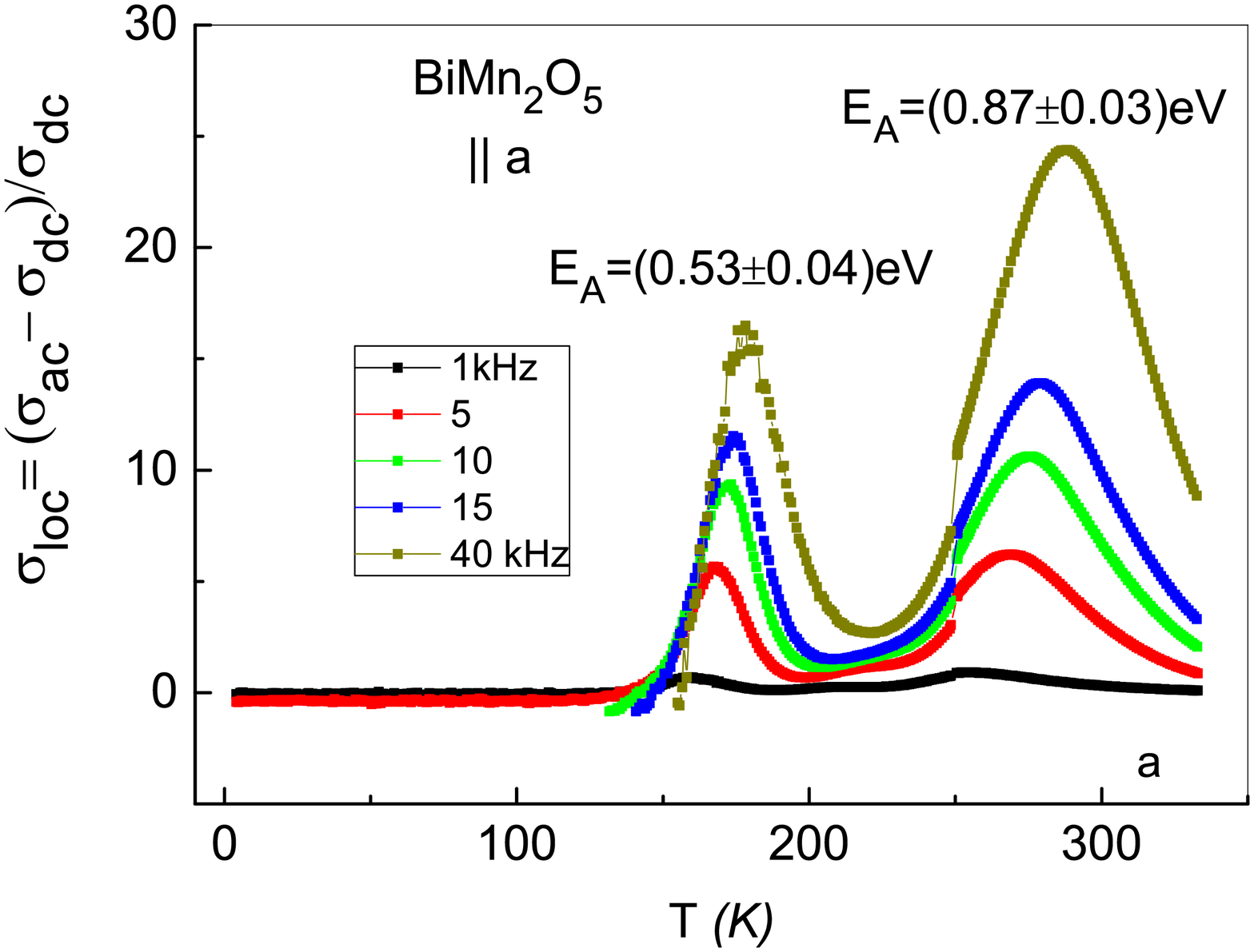}
  \includegraphics[width=0.45\textwidth, angle=0]{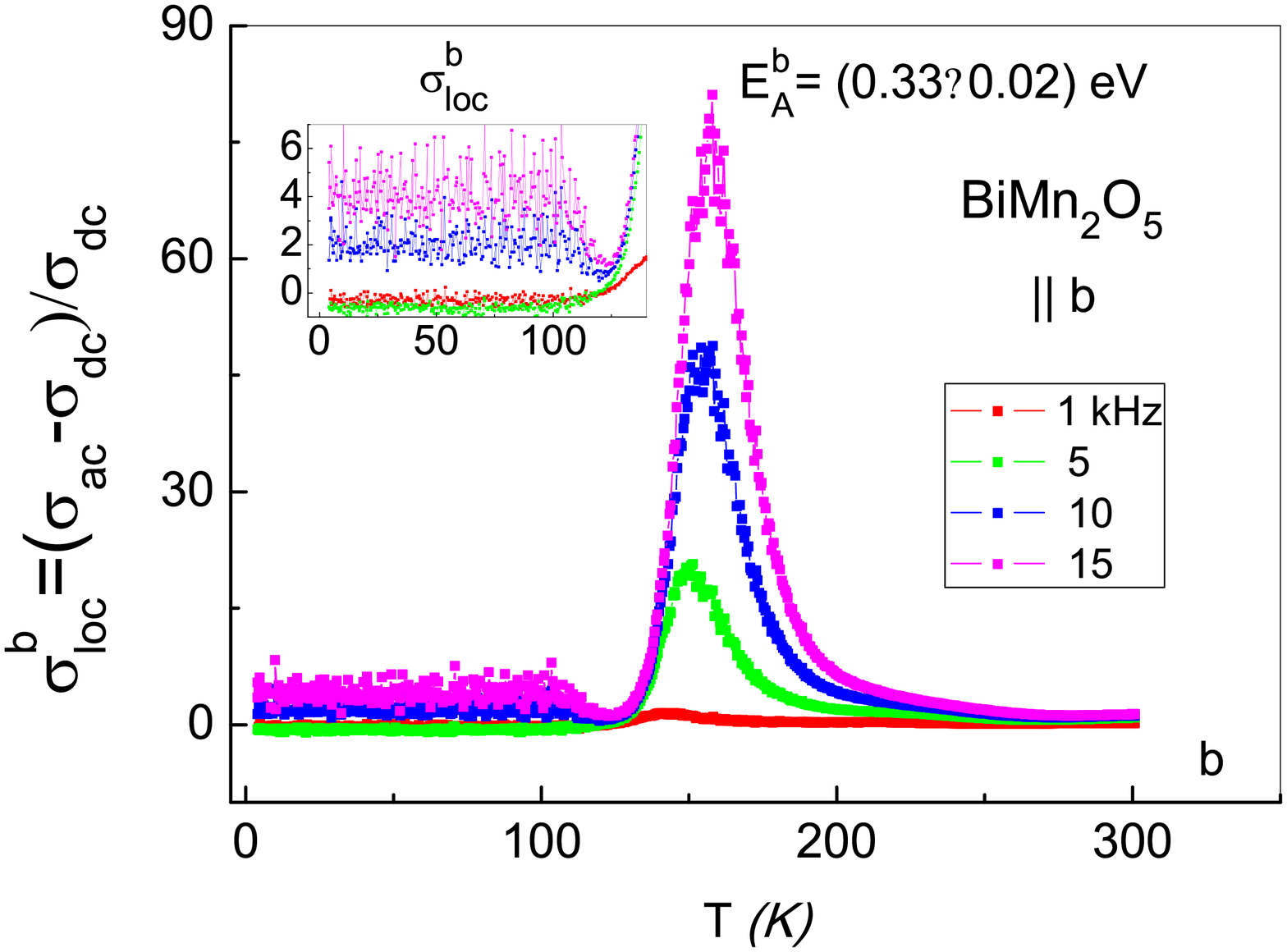}
  \includegraphics[width=0.45\textwidth, angle=0]{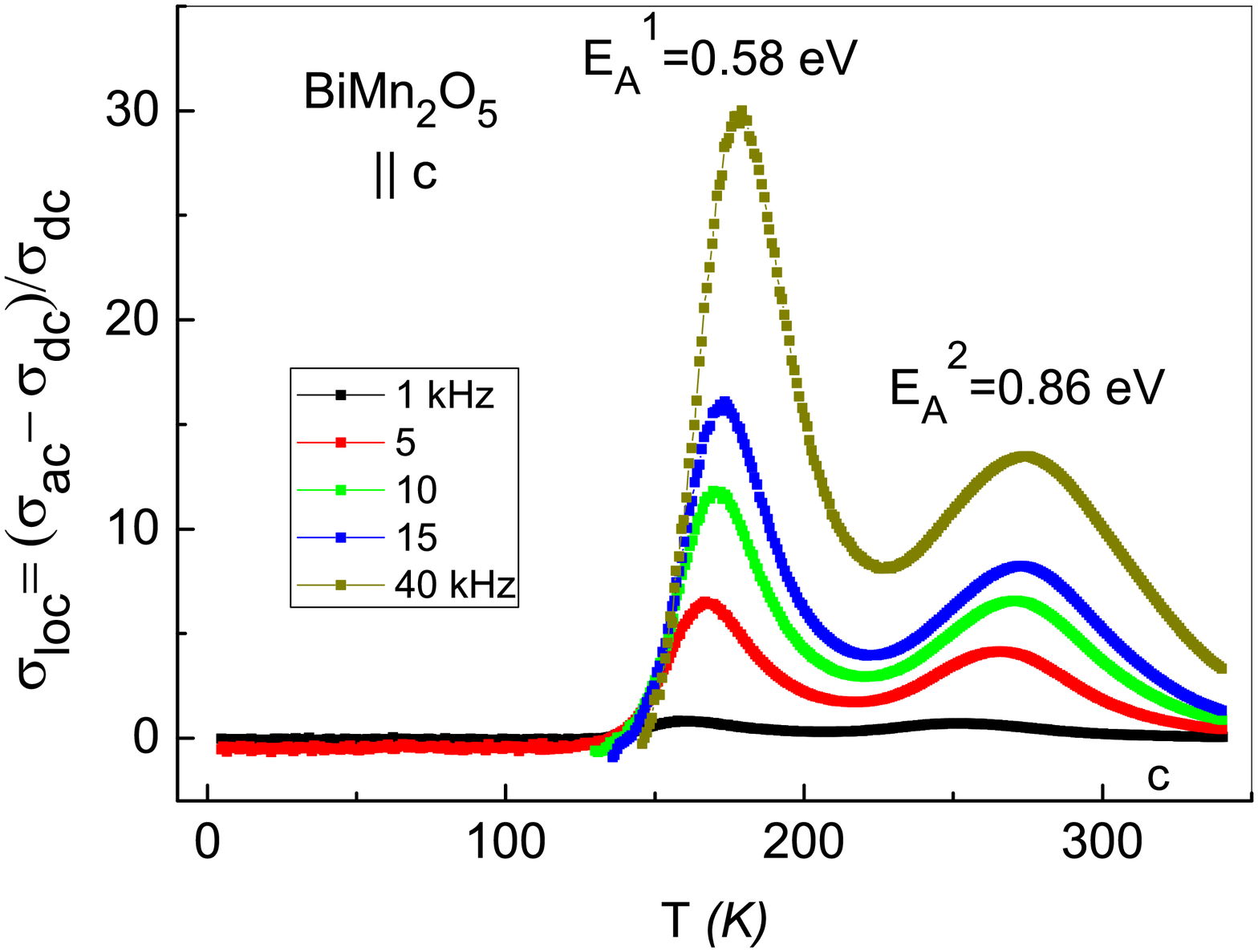}
  \caption{Temperature dependences of $\sigma_{loc}$ at different frequencies for BiMn$_2$O$_5$ along the $a$ axis (a),
   $b$ axis (b), and  $c$ axis (c). Inset shows the same dependences along the $b$ axis at $T < 125$ K at a larger scale.}
\label{fig7}
\end{figure}

\begin{figure}[htb]
  \includegraphics[width=0.45\textwidth, angle=0]{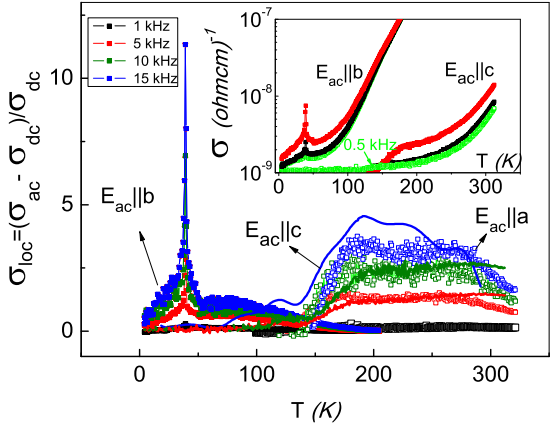}
    \caption{Temperature dependences of $\sigma_{loc}$ for GdMn$_2$O$_5$ along all axes (the $a$ axis --- unfilled symbols, the $b$ axis --- filled symbols,
the $c$ axis --- lines) for different frequencies. The inset shows temperature dependences of conductivity along the $b$ and $c$ axes for different frequencies.}
\label{fig8}
\end{figure}

Let us turn to a more detailed examination of the conductivity of the crystals. We deal with the real conductivity
$\sigma_1 = \omega\varepsilon^{''}\varepsilon_0$ \cite{Long} which is calculated from dielectric losses $\varepsilon^{''}$
($\omega$ is the angular frequency, $\varepsilon_0$ is the permittivity $\varepsilon^{\prime}$ at $\omega = 0$).
This conductivity depends on both the frequency and temperature. The low-frequency conductivities
are dispersion-free (percolation conductivity $\sigma_{dc}$).
The conductivity $\sigma_{ac}$ has a frequency dispersion: the higher the frequency, the higher the conductivity.
The frequency dispersion of this type
is typical of local conductivity (i.e., dielectric losses) in restricted domains ~\cite{Long}.
In our case, we attributed this local conductivity
to phase separation domains. The percolation conductivity (leakage) is attributable to the initial
crystal matrix. The relative local conductivity
$\sigma_{loc} = (\sigma_{ac} - \sigma_{dc})/\sigma_{dc}$ characterizes the ratio between
the phase separation domain local conductivity and matrix leakage.
Fig.~\ref{fig7}  shows $\sigma_{loc}$ for BiMn$_2$O$_5$ along the $a$, $b$, and $c$ axes (Figs.~\ref{fig7}a,
~\ref{fig7}b, and ~\ref{fig7}c, respectively). Along the $b$ axis only one, most intense, maximum
in $\sigma_{loc}$ in the temperature interval 125 -- 220 K is observed.
Below 100 K $\sigma_{loc}$ also exists, but at $T >100$ K it is transformed into leakage, and near 125 K
it again becomes a local conductivity (see the inset in Fig.~\ref{fig7}b). The activation barrier calculated
from the shifts of the local conductivity maxima
at $T > 125$ K (Fig.~\ref{fig7}b) exceeds $E_A^b$ derived from the $\varepsilon^{\prime}$ jumps only slightly
(Figs.~\ref{fig6}b and ~\ref{fig7}b).
At $T <100$ K the conductivities along the $a$ and $c$ axes were low and could not be measured correctly by our device.
There are two $\sigma_{loc}$ maxima along these axes at $T > 125$ K, the first is in the same temperature
interval 125 -- 220 K as along the $b$ axis; the second is in the temperature region 225 -- 350 K.
The presence of two maxima split in temperature along the $a$ and $c$ axes in BiMn$_2$O$_5$ means that the leakage sharply increases in the temperature interval between the maxima.
Such behaviors of $\sigma_{loc}$ along the $a$ and $c$ axes correlate with the temperature
behaviors of $\varepsilon^{\prime}$ along these axes (Fig.~\ref{fig4}). However, the activation barriers
calculated from the shifts of the first maxima of $\sigma_{loc}$ are almost twice as high
as the barriers derived from the $\varepsilon^{\prime}$ jumps (compare Fig.~\ref{fig6} and
Fig.~\ref{fig7}).
%
%
\noindent Tte inset in Fig.~\ref{fig8}
shows temperature dependences of conductivities of GdMn$_2$O$_5$ along the
$b$ and $c$ axes at different frequencies. The conductivity along the $a$ axis is close to that
along the $c$ axis. Of interest is the fact that the frequency-dependent conductivity jumps along
the $a$ and $c$ axes in GdMn$_2$O$_5$ also emerge near 125 K, like in BiMn$_2$O$_5$. The activation barrier $E_A$
in GdMn$_2$O$_5$ calculated from the conductivity jumps is $0.2$ eV. Fig.~\ref{fig8} shows
the temperature intervals in GdMn$_2$O$_5$ in which $\sigma_{loc}$ of the phase separation domains
exceeds considerably the matrix leakage.
Conductivity $\sigma_{loc}$ along the $b$ axis manifests itself up to $T\approx 125$ K.
At $T > 125$ K there is only the leakage which grows with temperature. At $T\simeq 125$ K
$\sigma_{loc}$ along the $a$ and $c$ axes rises abruptly and exists up to room temperature without a noticeable change.
The hysteresis loops and $P^{rem}$ along all axes in both crystals
are screened by leakage rather sharply at the temperatures $T^*$ at which
the leakage and local conductivities become comparable.
This means that the $T^*$ values correspond to the temperatures at which the potential barriers of the restricted
polar domain reorientations become equal to the kinetic energy of the itinerant electrons (leakage).

Differences between polar properties of phase separation domains in GdMn$_2$O$_5$ and BiMn$_2$O$_5$ are observed
along all axes of the crystals at $T >125$ K. At $T < 125$ K the ground states of phase separation
domains are similar for both crystals, which means that these states are formed inside the Mn subsystems
under the barriers of any origin. Near 125 K the leakage and local conductivities along the $b$ axes become comparable
in both crystals. The itinerant electrons which appear along the $b$ axes at $T > 125$ K
are localized anew in deeper potential wells thus passing into $\sigma_{loc}$ of the phase separation domains.
In GdMn$_2$O$_5$ these $\sigma_{loc}$ exceed leakage up to $T^*$ equal to 295 K and 325 K
along the $a$ and $c$ axes, respectively (see Fig.~\ref{fig5}).
In BiMn$_2$O$_5$ $\sigma_{loc}$ of phase separation domains emerges at $T > 125$ K along all crystal axes,
including the $b$ axis, due to the effect of Bi ions.
The localization of itinerant electrons in potential wells leads to deepening of these wells
and increases the activation barriers of phase separation domains. These $E_A$ barriers
can be calculated from the shifts of the temperature maxima of $\sigma_{loc}$
at different frequencies. Indeed, these barriers in BiMn$_2$O$_5$ turned out to be twice
as high as $E_A$ calculated from the $\varepsilon^{\prime}$ jumps. The latter barriers correspond to the
condition when the kinetic energy of the itinerant electrons emerging along the $b$ axis becomes
comparable to $E_A$ calculated from the $\varepsilon^{\prime}$ jumps.
The values of $T^*$ in BiMn$_2$O$_5$ along different axes correlate with the temperatures
at which the conductivity at the lowest frequency
(leakage) begins to increase sharply in the first $\sigma_{loc}$ maxima
(compare Fig.~\ref{fig4} and Fig~\ref{fig7}).
%
%
\noindent  Note that near $T_C\approx 30$ K dispersion-free anomalies typical of the ferroelectric phase transition
manifest themselves in GdMn$_2$O$_5$ on the background of $\sigma_{loc}$ and $\sigma_{ac}$ along the $b$ axis
(see Fig.~\ref{fig8}). This is due to the fact that a maximum
in $\varepsilon^{''}$ should be observed near $T_C$, while in BiMn$_2$O$_5$ a low polarization
$P_b$ does not manifest itself on the background of $\sigma_{loc}$ (Fig.~\ref{fig7}b).

Both the intrinsic polarization and $\sigma_{loc}$ of restricted polar
phase separation domains give contributions into the P1 and N1 curves in measurements
of hysteresis loops for GdMn$_2$O$_5$ and BiMn$_2$O$_5$
at $T < T^*$ by the PUND method, while $\sigma_{loc}$ contributes significantly to the P2 and N2 curves
in the same temperature interval. At these temperatures $P^{rem}$ emerges in these crystals.
Near $T^*$ at which $P^{rem}$ tends to zero, the polarization and local conductivity contributions
to the P1-P2 and N1-N2 curves start to decrease simultaneously. At the temperatures at which $P^{rem}=0$,
these contributions are transformed into linear dependences on E.
At $T > T^*$ the loop restores as an inverted loop,
and the leakage contribution begins to dominate in it.

\subsection{X-ray high-sensitivity diffraction study}
\label{X-ray}

\begin{figure}[htb]
  \includegraphics[width=0.45\textwidth,angle=0]{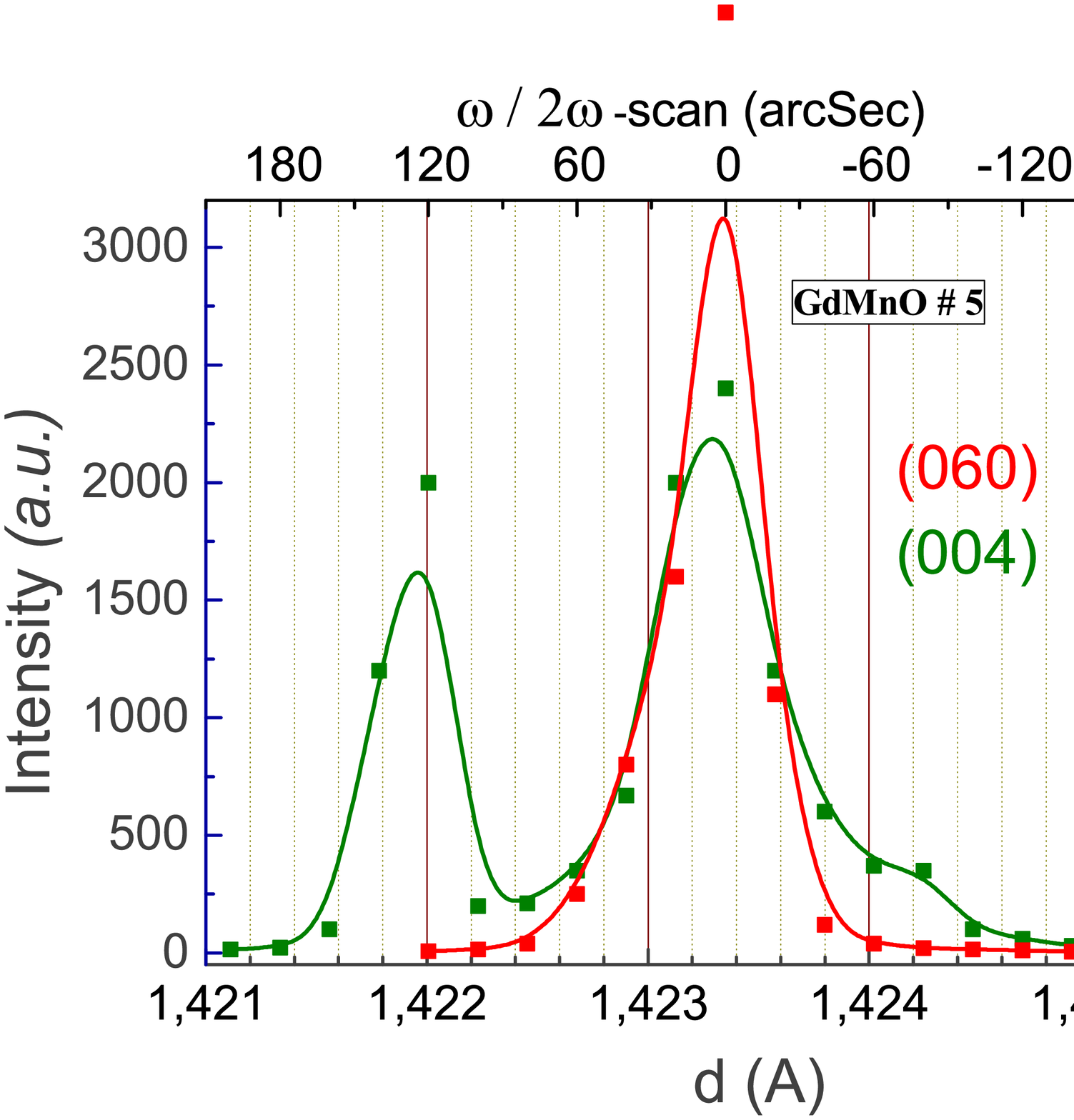}
  \includegraphics[width=0.45\textwidth,angle=0]{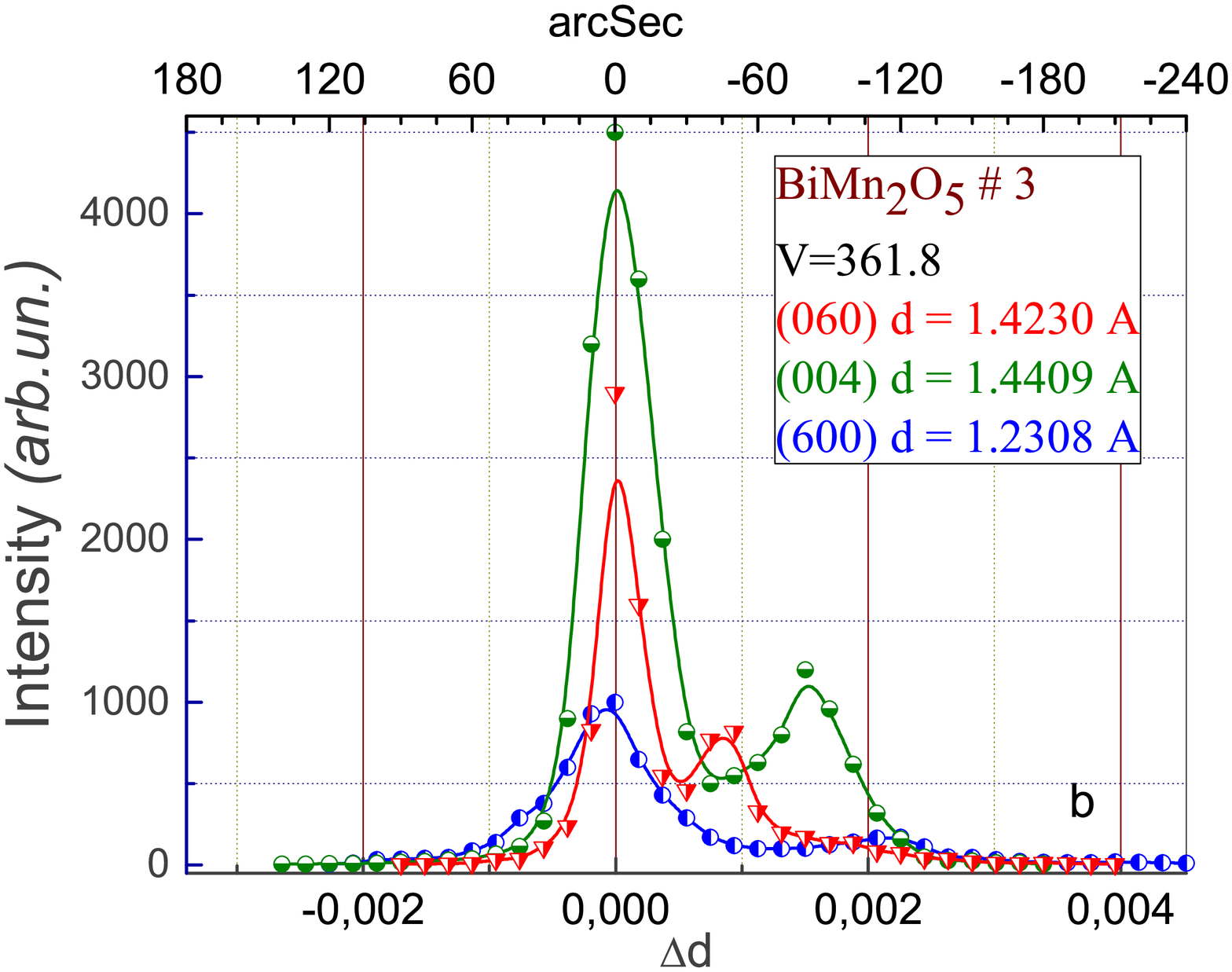}
\caption{Angular intensity distributions of ~{\bf (600)},
~{\bf (060)}, and {\bf (004)}$_{CuK\alpha 1}$ Bragg reflections as functions of interplanar distances (d $\AA$) in GdMn$_2$O$_5$ (a) and BiMn$_2$O$_5$ (b).
Parameters of GdMn$_2$O$_5$ lattice are $a = 7,3568\pm 0.0002$~\AA; $b = 8,5398\pm 0.0002$~\AA; $c = 5,6920\pm 0.0002$~\AA.}
\label{fig9}
\end{figure}

The experimental evidence of the existence of restricted polar phase separation domains of GdMn$_2$O$_5$ and
BiMn$_2$O$_5$ was obtained in the X-ray high-sensitivity diffraction study carried out
at room temperature (see Figs.~\ref{fig9}a and ~\ref{fig9}b, respectively).
The angular intensity distributions of {\bf (004)}, {\bf (060)} and {\bf (600)}$_{CuK \alpha 1}$
Bragg reflections were detected in the 3-crystal
regime with the $\theta/2\theta$ scan.
As a monochromator and an analyzer, germanium crystals in the {\bf (004)} reflection were used, which allowed
conditions of nearly dispersion-free high-resolution ($\sim 2''$) survey geometry to be realized.
Fig.~\ref{fig9}a shows a single diffraction maximum of {\bf (060)} and two diffraction
maxima of the {\bf (004)} Bragg reflections recorded from different planes of a single crystal of GdMn$_2$O$_5$.
These planes are perpendicular to the $b$ and $c$ axes, respectively. The {\bf (004)} Bragg reflection
positions are characterized by slightly different interplanar spacings
$d$ ($\Delta d\approx 0.0015$ \AA). These {\bf (004)} reflections, which have comparable intensities and
half-widths ($\sim 20''$) clearly point to the coexistence of two phases with slightly
differing $c$ lattice parameters which differ in the third decimal place.
The Bragg peak along the $b$ axis with a similar half-width is not split, i.e., it is
identical to these two phases. The positions of all Bragg peaks
nearly coincide with those for GdMn$_2$O$_5$ with the generally accepted
{\it Pbam} symmetry. This means that two phases we detected are high-quality GdMn$_2$O$_5$
single crystal phases with similar large correlation
lengths $R_c$. The phase with a higher intensity of the {\bf (004)} Bragg peak can be attributed
to the original matrix. The phase with a lower intensity of such a peak can be attributed to the phase
separation domains. A similar picture of the angular intensity distributions
of the {\bf (004)}, {\bf (060)} and {\bf (600)}$_{CuK \alpha 1}$ Bragg reflections is also
observed for BiMn$_2$O$_5$, but splitting into two peaks
for the reflexes along all axes of the crystal is detected (Fig.~\ref{fig9}b). It is natural to attribute this
to lattice distortions caused by Bi ions
along all crystal axes. Note that the structural quality of both crystals is similar.
%
%

\subsection{Duscusion}
\label{discusion}

Let us discuss the shape and structure of phase separation domains in GMn$_2$O$_5$ and BiMn$_2$O$_5$.
As noted above, the phase separation domains in the form of $1D$ superlattices were observed below the temperatures
of magnetic and ferroelectric orderings in a number of RMn$_2$O$_5$ with various $R$ ions, including $R = Gd$ and $Bi$
~\cite{JETPLett2012,JPCM2012,JPCS2014}. A set of ferromagnetic resonances was detected from
individual layers of superlattices. The intensity of these resonances sharply decreased when  $T\simeq 30 - 40$ K was approached.
Since the phase separation domains should exist at all temperatures, this fact means that the structure of the phase separation
domains must change at $T > 40$ K. We manage to study the temperature evolution of feromagnetic resonances of  layered superlattices at $T > 40$  K only in Eu$_{0.8}$Ce$_{0.2}$Mn$_2$O$_5$, in which  the phase separation domains were similar to those in EuMn$_2$O$_5$,  but their
concentration was much high ~\cite{PRB2009,JPCM2011}. In temperature interval $100 K > T > 40$  K only one ferromagnetic resonance
signal was observed in Eu$_{0.8}$Ce$_{0.2}$Mn$_2$O$_5$.
This resonance corresponded to the low-temperature superlattice
layer having the maximal barrier and conductivity  ~\cite{JPCM2012}.
Near $100$ K an intensity of this resonance became indistinguishable due to growth of conductivity.
At room temperature the structure of phase separation domains was installed in Eu$_{0.8}$Ce$_{0.2}$Mn$_2$O$_5$ in X-ray diffraction study ~\cite{PRB2009}. The two $(004)$  well defined Bragg peaks were accompanied
by the characteristic oscillation peaks independent on the interplanar spacings $d$, which pointed  to existence
of  the layered superstructure in this crystal. As shown in ~\cite{PRB2009}, the layered superstructure consisting
of the alternating original crystal EuMn$_2$O$_5$ layers and conductive layers of phase separation
emerged in Eu$_{0.8}$Ce$_{0.2}$Mn$_2$O$_5$ at all temperatures above 40 K.  No similar oscillation peaks on the  background
of the Bragg reflexes were observed in X-ray studies of GdMn$_2$O$_5$ and BiMn$_2$O$_5$ at room temperature (Fig.~\ref{fig9}),
i.e., there were no  layered superstructures in these crystals. Most likely the phase separation domains in GdMn$_2$O$_5$ and
BiMn$_2$O$_5$ had the shape of nanoscale droplets at $T > 40$ K.
The possibility of different geometries of the phase separation domains in Coulomb-frustrated systems,
including manganites and high-temperature superconductors, was shown theoretically in  ~\cite{OrtixPRL,OrtixJPCM,Kugel2009}.
The properties of the phase separation domains in the form of nanoscale droplets for manganites were discussed in
\cite{Gorkov,Kagan}.

Now consider in more detail what are polar phase separation domains. There is a probability that $e_g$ electrons of
some Mn$^{3+}$ ions tunnel to Mn$^{4+}$ ions in the original RMn$_2$O$_5$ matrix. These valence electrons
and recharged Mn$^{3+}$-Mn$^{4+}$ ion pairs are accumulated
in the restricted phase separation domains inside the original matrix because phase separation
is energetically favorable ~\cite{Gorkov,Kagan}.
Since Mn$^{3+}$ ions appear in the Mn$^{4+}$ ion positions (i.e., in the octahedral oxygen surrounding),
they become Jahn-Teller ions and give rise to
local deformations of these octahedra. In turn, Mn$^{4+}$ ions appear in the Mn$^{3+}$ ion positions
(in noncentrosymmetrical pentagonal pyramids) and local distortions arise near these ions as well.
As a result, structural distortions caused by both these factors occur inside restricted phase separation domains.
We believe that phase separation domains are noncentrosymmetric, and their sizes are large enough
to give rise to ferroelectric single-domain states inside them.

The restricted polar phase separation domain states existing in GdMn$_2$O$_5$ and BiMn$_2$O$_5$ original matrices
are analogs of the superparaelectric state formed by an ensemble of spherical ferroelectric nanoparticles
in a dielectric matrix theoretically studied in ~\cite{Glinchuk}. It was shown in ~\cite{Glinchuk} that at low temperatures a homogeneous polarization could exist in these particles
(in our case restricted polar phase separation domains) if their sizes $R$ were less than the correlation
radius $R_c$ but larger than the critical radius $R_{cr}$ of the size-driven ferroelectric-paraelectric
phase transition. Under these conditions all dipole moments inside particles are aligned due to correlation effects.
Surface screening of depolarization fields makes single-domain states energetically favorable.
If $R < R_{cr}$, separate paraelectric dipoles are uncorrelated and represent local polar defects
which can only increase the width of the Bragg peak original matrix. The fact that the well-defined
Bragg reflections related to the restricted polar phase
separation domains are observed in our study together with similar Bragg peaks of the original matrix
indicates that the conditions for the ferroelectric single-domain states of phase separation regions
given in ~\cite{Glinchuk} are fulfilled.  It was also found in ~\cite{Glinchuk}
that a frozen superparaelectric phase could emerge for an ensemble of ferroelectric nanoparticles
in a dielectric matrix. In this phase remanent polarization
and hysteresis loop arose at the temperatures lower than the freezing temperature $T_f$.
$T_f$ was defined from the condition that the potential barriers
of nanoparticle polarization reorientation became equal to the thermal activation energy $\sim k_BT$.
At $T > T_f$ the conventional superparaelectric state occurred for which
there are no hysteresis loops and remanent polarization.
It was also accepted in \cite{Glinchuk} that the temperature of thermal destruction of intrinsic
nanoparticle ferroelectricity was $T_{cr}\gg T_f$.
So, the temperature behavior of the GdMn$_2$O$_5$ and BiMn$_2$O$_5$ polarization is also
analogous to that of the frozen superparaelectric state discussed in ~\cite{Glinchuk}.
$T^*$ considered above can be regarded as $T_f$. As noted above, the $T^*$ values correspond
to the temperatures at which the potential barriers of the restricted
polar domain reorientations become equal to the kinetic energy of the itinerant electrons (leakage).
The restricted phase separation domains still exist at $T > T^*$ due to energetic favorability of
the phase separation caused by strong interactions. But the frozen superparaelectric state
turns into the conventional superparaelectric one near $T^*$.

Since $P-E$ hysteresis loops are measured under the field E applied along different axes, electric polarizations
inside phase separation domains are induced along these axes. Indeed, the application of field E along
any crystal axis initiates a drift of valence $e_g$ electrons localized inside phase separation
domains in this direction. These electrons recharge $Mn^{3+}$ and $Mn^{4+}$ ions inside the phase separation
domains. As a result, the spatial distribution of the $Mn^{3+}$ and $Mn^{4+}$ ions and structural distortions
inside phase separation domains are bound to change, thereby giving rise
to the polarization along the E direction. Thus, the actual symmetry of RMn$_2$O$_5$ at room temperature
can be established only in polarization measurements
in $E=0$. We believe that the additional structural reflexes typical of the noncentral monoclinic
structure reported in ~\cite{PRL2015}
can be related to the frozen superparaelectric state arising in RMn$_2$O$_5$ at $T\leq T^*$. Note that in the cases when temperatures $T^*$ are sufficiently high, the correlations between
the phase separation domains can occur in the entire crystal volume. This situation was observed earlier
in Eu$_{0.8}$Ce$_{0.2}$Mn$_2$O$_5$  at room temperature ~\cite{PRB2009}.

As noted above, phase separation and charge carrier self-organization give rise to a dynamic equilibrium
of the phase separation domain states with a balance
between attraction (double exchange, Jahn-Teller effect) and Coulomb repulsion of charge carriers ~\cite{Gorkov,Kagan,PRB2009}. The formation of the phase separation domains due to a balance
between strong interactions leads to specific features in RMn$_2$O$_5$ properties.
First, polar phase separation domains are bound to emerge up to high temperatures,
thus giving rise to high-temperature polarization.
Second, changes in polar phase separation domains under varying E rapidly relax to the dynamic
equilibrium states after E is switched off.
These features manifested themselves in our experiments and necessitated the use of the PUND method for measuring the polarization hysteresis loops
which were discussed above.

The application of magnetic field H increases the barriers at the phase separation domain boundaries due to
the double exchange growth, thus increasing the $T^*$ temperatures in GdMn$_2$O$_5$
(see the right panels in Fig.~\ref{fig5}). The field H oriented along the $b$ axis in GdMn$_2$O$_5$
also enhances the polarization induced by the restricted polar domains due to the increasing probability
of charge transfer between $Mn^{3+}- Mn^{4+}$ ion pairs with the greatest distance between them
(see the caption to Fig.~\ref{fig9}a).
It was found that application of magnetic field $H = 6$ T to BiMn$_2$O$_5$ had a much weaker effect
on phase separation domain properties and polarization
than in GdMn$_2$O$_5$ because of strong structural distortions of the crystal field by Bi ions.

\section{Conclusion}
\label{conclusion}

Thus, remanent polarizations and hysteresis loops originating from the frozen superparaelectric state
of similar ferroelectric restricted polar phase separation domains arranged inside the initial matrix
of RMn$_2$O$_5$ ($R = Gd$ and $Bi$) have been revealed in wide temperature intervals from 5 K up to $T^*$
depending on the crystal axis. The restricted polar phase separation domains emerge due to phase
separation and charge carrier self-organization. At $T\simeq T^*$ the potential barriers of
the restricted polar domain reorientations become equal to the kinetic energy of the itinerant electrons (leakage).
In GdMn$_2$O$_5$, the polarization of this type emerges along the $c$ axis up to $T^*\simeq 325$ K.
In BiMn$_2$O$_5$, the maximum temperature $T^*\simeq 220$ K is observed along the $b$ axis.
At $T\geq T^*$ the frozen superparaelectric state turns into the conventional superparaelectric one.
The polarizations observed can be attributed to the magnetically induced polarizations
since the magnetic double exchange between Mn$^{3+}$-Mn$^{4+}$ ion pairs is the key interaction
giving rise to the phase separation domain formation.
The effect of magnetic field on these polarizations in GdMn$_2$O$_5$ demonstrates that the magnetoelectric
coupling exists in the paramagnetic phase as well. Application of electric field E modifies the structure and shape
of the phase separation domains, giving rise to the polarization along the E direction.
Additional structural reflexes typical of the noncentral monoclinic structure reported in ~\cite{PRL2015} can be related to the frozen
superparaelectric state arising in RMn$_2$O$_5$ at $T\leq T^*$.
In the cases when temperatures $T^*$ are sufficiently high, the correlations between the phase separation
domains can emerge in the entire crystal volume thus giving rise the noncentosymmetricity of the whole crystals.

\section{Acknowledgments}
The work was supported by the Government of Russian Federation (project N0.14.B25.31.0025).

\end{document}